\documentclass{aa}

\usepackage{graphicx}
\usepackage{txfonts}
\usepackage{hyperref}
\usepackage[caption=false, labelformat=empty]{subfig}
\usepackage{xcolor}
\usepackage{lscape}
\usepackage{placeins}

\usepackage{orcidlink}
\newcommand\orcidicon[1]{\orcidlink{#1}}

\begin{document}

\title{BASS LI. Cool gas supply of H$\textrm{I}$-massive local Seyfert galaxies}

\author{
Jeein Kim\inst{1,2\orcidicon{0000-0002-3170-7434}}\fnmsep\thanks{jikim.astro@gmail.com}
\and
Aeree Chung\inst{1\orcidicon{0000-0003-1440-8552}}\fnmsep\thanks{Corresponding author: achung@yonsei.ac.kr}
\and
O. Ivy Wong\inst{3,4\orcidicon{0000-0003-4264-3509}}
\and
Junhyun Baek\inst{5,1\orcidicon{0000-0002-3744-6714}}
\and
Chandrashekar Murugeshan\inst{6,3\orcidicon{0000-0002-4366-868X}}
\and
Michael J. Koss\inst{7\orcidicon{0000-0002-7998-9581}}
\and
Kyuseok Oh\inst{5\orcidicon{0000-0002-5037-951X}}
\and
Franz Erik Bauer\inst{8\orcidicon{0000-0002-8686-8737}}
\and
Chin-Shin Chang\inst{9\orcidicon{0000-0001-9910-3234}}
\and
Yaherlyn Diaz\inst{10\orcidicon{0000-0002-8604-1158}}
\and
Kohei Ichikawa\inst{11,12\orcidicon{0000-0002-4377-903X}}
\and
Darshan Kakkad\inst{13\orcidicon{0000-0002-2603-2639}}
\and
Minjin Kim\inst{1,14\orcidicon{0000-0002-3560-0781}}
\and
Macon Magno\inst{15,3\orcidicon{0000-0002-1292-1451}}
\and
Ignacio del Moral-Castro\inst{16\orcidicon{0000-0001-8931-1152}}
\and
Richard Mushotzky\inst{17\orcidicon{0000-0002-7962-5446}}
\and
Alessandro Peca\inst{7,18\orcidicon{0000-0003-2196-3298}}
\and
Claudio Ricci\inst{19,10,20\orcidicon{0000-0001-5231-2645}}
\and
Matilde Signorini\inst{21,22,23\orcidicon{0000-0002-8177-6905}}
\and
Miguel Parra Tello\inst{16\orcidicon{0000-0001-5649-7798}}
\and
Benny Trakhtenbrot\inst{24\orcidicon{0000-0002-3683-7297}}
\and
Jong-Hak Woo\inst{25\orcidicon{0000-0002-8055-5465}}
}

\institute{Department of Astronomy, Yonsei University, 50 Yonsei-ro, Seodaemun-gu, Seoul 03722, Republic of Korea
\and
National Radio Astronomy Observatory, 1011 Lopezville Rd, Socorro, NM 87801, USA
\and
ATNF, CSIRO, Space and Astronomy, PO Box 1130, Bentley, WA 6102, Australia
\and
ICRAR, University of Western Australia, 35 Stirling Highway, Crawley, WA 6009, Australia
\and
Korea Astronomy and Space Science Institute, 776 Daedeok-daero, Yuseong-gu, Daejeon 34055, Republic of Korea
\and
Australian SKA Regional Centre, University of Western Australia, 35 Stirling Highway, Crawley WA 6009, Australia
\and
Eureka Scientific, 2452 Delmer Street, Suite 100, Oakland, CA 94602-3017, USA
\and
Instituto de Alta Investigaci{\'{o}}n, Universidad de Tarapac{\'{a}}, Casilla 7D, Arica, Chile
\and
Department of Astronomy, University of Geneva, Chemin Pegasi 51, 1290 Versoix, Switzerland
\and
Instituto de Estudios Astrof\'isicos, Facultad de Ingenier\'ia y Ciencias, Universidad Diego Portales, Av. Ej\'ercito Libertador 441, Santiago, Chile
\and
Frontier Research Institute for Interdisciplinary Sciences, Tohoku University, Sendai, Miyagi 980-8578, Japan
\and
Astronomical Institute, Tohoku University, Aramaki, Aoba-ku, Sendai, Miyagi 980-8578, Japan
\and
Centre for Astrophysics Research, Department of Physics, Astronomy and Mathematics, University of Hertfordshire, Hatfield AL10 9AB, UK
\and
Department of Astronomy and Atmospheric Sciences, Kyungpook National University, Daegu 41566, Republic of Korea
\and
George P. and Cynthia Woods Mitchell Institute for Fundamental Physics and Astronomy, Texas A\&M University, College Station, TX 77845, USA
\and
Instituto de Astrof\'isica, Facultad de F\'isica, Pontificia Universidad Cat\'olica de Chile, Casilla 306, Santiago 22, Chile
\and
Department of Astronomy, University of Maryland, College Park, MD 20742, USA
\and
Department of Physics, Yale University, P.O. Box 208120, New Haven, CT 06520, USA
\and
Department of Astronomy, University of Geneva, ch. d'Ecogia 16, 1290, Versoix, Switzerland
\and
Kavli Institute for Astronomy and Astrophysics, Peking University, Beijing 100871, China
\and
Dipartimento di Matematica e Fisica, Università degli Studi di Roma 3, Via della Vasca Navale, 84, 00146 Roma, Italy
\and
INAF – Osservatorio Astrofisico di Arcetri, Largo Enrico Fermi 5, I-50125 Firenze, Italy
\and
European Space Agency, ESTEC, Keplerlaan 1, 2201 AZ Noordwijk, the Netherlands
\and
School of Physics and Astronomy, Tel Aviv University, Tel Aviv 69978, Israel
\and
SNU Astronomy Research Center, Seoul National University, 1 Gwanak-ro, Gwanak-gu, Seoul 08826, Republic of Korea
}

\date{Received 19 August 2025 / Accepted 24 February 2026}

\abstract 
{We present neutral atomic hydrogen (H{\scriptsize I}) imaging observations of 22 H{\scriptsize I}-rich ($M_{\rm HI}\gtrsim 10^{9.7}M_\odot$), hard X-ray-selected local Seyferts to explore how cool gas is supplied to active galactic nuclei (AGN) hosts. 
The sample predominantly resides in group-like, gas-rich environments. About 80\% (18/22) of the galaxies have H{\scriptsize I}-detected neighbors, 61\% (11/18) of which clearly exhibit strong lopsidedness, one-sided gas tails, and/or gas structures connecting to nearby companion galaxies, suggesting gas exchange histories.
We examine the H{\scriptsize I} size-mass relation and star formation properties of these H{\scriptsize I}-rich AGN hosts, finding no systematic deviations from known scaling relations. 
In most cases, our samples are the most massive systems within their respective groups, implying that our sample is more likely to acquire gas rather than lose it.
Interestingly, galaxies with more extended H{\scriptsize I} disks show stronger AGN activity. 
Considering that extended H{\scriptsize I} is often associated with external processes, this finding suggests that environmentally accreted gas - through galaxy interactions and gas exchange with neighboring systems - may have played a role in supplying additional fuel to the AGNs in our sample.
Notably, the H{\scriptsize I} extent–AGN activity correlation becomes even tighter for those AGN hosts whose neighboring galaxies are gas poor or lack H{\scriptsize I}, further supporting externally supplied gas as a fuel source.
}

   \keywords{Galaxies: active -- Galaxies: evolution -- Galaxies: nuclei -- Galaxies: ISM -- Galaxies: Seyfert}

   \maketitle

\newcommand{\HI}{\mbox{\sc      H{i}}}
\newcommand{\Hmol}{\mbox{\sc  H$_{2}$}}
\newcommand{\barolo}{\textsc{${\rm 3D}$-barolo }}
\newcommand{\sofia}{\textsc{SoFiA-2 }}
\newcommand{\barolonogap}{\textsc{{${\rm 3D}$-barolo}}}
\newcommand{\sofianogap}{\textsc{SoFiA-2}}
\newcommand{\HIX}{\mbox{\sc      H{ix}}}

\section{Introduction} \label{sec:intro1}

Current models of galaxy evolution suggest that the quenching of star formation can be regulated by the accretion-driven growth of supermassive black holes \citep[SMBHs; e.g.,][]{Croton2006, Vogelsberger2014, Schaye2015}. To understand the relationship between active galactic nuclei (AGNs) and star formation, and hence galaxy evolution, it is essential to examine the cold gas, particularly the atomic hydrogen (\HI) gas reservoir ($\sim\text{10}^2-\text{10}^4$K), as it supplies the material for molecular clouds and stars \citep[e.g.,][]{Leroy2008, Bigiel2010}. In particular, the properties of \HI\ gas in AGN hosts can provide valuable insights into gas accretion and removal mechanisms, serving as a sensitive tracer of interactions between other galaxies and their surroundings.

Gas inflows in the form of cold gas accretion from the circumnuclear environment are needed to activate an SMBH. Luminous AGNs require large amounts of gas to be rapidly funneled toward the SMBH, whereas low-luminosity AGNs can sustain activity through stellar mass loss \citep[e.g.,][]{StorchiBergmann_2019}. Intriguingly, observational studies of X-ray-selected AGNs, which typically target more luminous AGNs in a relatively unbiased manner, find that they preferentially populate denser environments \citep[e.g.,][]{Koss2010, Starikova2011, Cappelluti2012, Powell2018}.
This trend underscores the importance of investigating the surroundings of AGN hosts to better understand how they are supplied with gas fuel.

\HI\ gas can be an important probe not only of the molecular gas reservoir, and hence star formation, but also of how galaxies interact with their environments. Its typically large extent and low density make it vulnerable to the surroundings, leaving useful footprints (e.g., \HI\ bridges, tidal tails) to study interaction histories that are not visible at other wavelengths \citep[e.g.,][]{Hibbard1996, Zuo2022, Tillman2023}. In addition, \HI\ absorption can be a direct tracer of cold gas inflows and outflows near AGNs \citep[e.g.,][]{Morganti2013c_Science, Woo2016, Allison2019, Baek2022}. It has been shown that the distribution of \HI-absorbing gas is closely correlated with the distribution of soft X-ray absorbing media \citep{Moss2017}. However, absorption studies are limited to relatively cold gas components only along a limited line of sight to background continuum sources that are generally compact or unresolved, and thus imaging \HI\ emission is essential to obtain more complete pictures.

Due to the high cost of imaging with interferometers, previous \HI\ studies of relatively large AGN samples have often relied on single-dish radio telescopes. However, the connection of \HI\ gas to the AGN host galaxies from single-dish observations or stacking experiments remains inconclusive. Some studies find similar \HI\ gas fractions as non-AGN samples \citep[e.g.,][]{Fabello2011b, Gereb2015b, Bradford2018, Ellison2019}. In contrast, some find increased \HI\ fractions \citep[e.g.,][]{Ho2008_ApJ, Berg2018}. Integrated \HI\ mass measurements or gas fractions alone, therefore, seem insufficient to probe how \HI\ gas is accreted or removed from galaxies. A comprehensive \HI\ mapping study is therefore crucial for linking the AGN and \HI\ gas, i.e., how the cold gas is fed to and removed from the host galaxies.

There have been some efforts to image AGN hosts in \HI. For example, \cite{Haan2008} compared the resolved \HI\ properties of Seyfert galaxies with those of low-ionization nuclear emission-line regions (LINERs), finding that companions and/or disturbed features in the outer \HI\ disk are not rare regardless of the AGN type. \cite{Kuo2008} also found that most of their AGN samples show tidally disturbed \HI\ gas morphologies compared to a control sample \citep{Tang2008}. They also found that the majority ($\sim$78\%) of AGN hosts have neighboring galaxies within a projected distance of $\leq$100 kpc.

To systematically investigate the various gas fueling mechanisms associated with AGN hosts, we have initiated an \HI\ imaging study of approximately 100 hard X-ray-selected AGNs detected in atomic hydrogen as part of the BASS survey. To achieve this, we have compiled archival data and conducted our observations of targets lacking \HI\ images or requiring deeper imaging.

The mapping of \HI\ in a hard X-ray-selected sample has not been made before collectively. The hard X-ray selection is an important distinction from previous studies because hard X-ray emission yields a more complete census including obscured AGNs, which can be linked to gas-rich or early merger phases \citep[][]{Blecha2018MNRAS.478, KimMJ2021ApJS}. While soft X-rays originate from the vicinity of the accretion disk, the hard X-ray emission is emitted from a region much closer to the central black hole \citep[][]{Petrucci2013}. The strength of interferometric \HI\ observations is the ability for \HI\ to trace potential interactions with an AGN host’s local environment. Therefore, while previous studies of gas and AGN have been able to connect the AGN state to the star formation fuel that is the gas \citep[e.g.,][]{Koss2021}, thus far, the connection between environment interactions and quasar activity is still an open area of investigation.

One of the key questions we aim to address with the hard X-ray–selected AGN sample is how frequently their host galaxies experience externally driven gas accretion, and whether such large-scale processes have any impact on the central AGN activity. This study therefore focuses on the 22 most \HI-rich systems within the $\sim$100 BASS-\HI\ galaxies, as these are the cases most likely to show evidence of recent or ongoing gas accretion. Although the sample size in this paper happens to be comparable to previous \HI\ mapping surveys \citep[e.g.,][]{Kuo2008, Haan2008}, our hard X-ray–selected sample benefits from comprehensive AGN-related measurements, allowing us to examine how galaxy-scale properties relate to those of the central AGNs. The \HI\ data for the full sample are currently being processed and will be presented in a separate paper, enabling a more statistically robust analysis of the \HI\ properties of X-ray–selected AGN hosts.

The structure of this paper is as follows: In Section \ref{sec:obsdata}, we introduce the selected sample, followed by a description of the observational parameters and data reduction procedures. In Section \ref{sec: results}, we describe the \HI\ characteristics of individual galaxies, how \HI-related quantities are derived, and investigate various global properties and scaling relations such as the \HI\ size-mass relation and star formation properties. In Section \ref{sec:discuss_sec}, we discuss their local environments and the AGN feeding mechanisms of our sample. Finally, we present our conclusions in Section \ref{sec: conclusion}. Throughout this paper, we adopt the cosmological parameters of $H_0 \text{= 70 km}\;\text{s}^{-1}\;\text{Mpc}^{-1}$, $\Omega_{M}$ = 0.3, and $\Omega_\Lambda$ = 0.7.

\vspace{5mm}

\section{Observations and data reduction} \label{sec:obsdata}

\subsection{The sample}
\begin{table*}
\caption{BASS DR2 information for this study. \label{tab:bass-hi-rich-sample-bass}}
\centering
\begin{tabular}{cllccccccc}
\hline\hline
BAT ID & Galaxy Name & Alternative Name & Type & $z$ & Dist & $\log M_\mathrm{BH}$ & $\log {L}_{\rm bol}$ & $\log N_{\rm HI}$ & $\log \lambda_{\rm Edd}$ \\
& & & & & (Mpc) & $(M_\odot)$ & $\log({\rm erg} \;{\rm s}^{-1})$ & (cm$^{-2}$) & \\
\hline
33 & \text{NGC262} & \text{Mrk348} & \text{Sy1.9} & 0.01467 & 63.5 & 6.80 & 44.73 & 23.12 & -0.25 \\
64 & \text{NGC452} & \text{UGC820} & \text{Sy2} & 0.01690 & 73.3 & 8.16 & 43.68 & 22.46 & -2.65 \\
83 & \text{ESO353-G009} & \text{MCG-6-4-43} & \text{Sy2} & 0.01639 & 71.1 & 7.96 & 43.82 & 23.90 & -2.32 \\
96 & \text{MCG-01-05-047} & \text{ } & \text{Sy2} & 0.01673 & 72.6 & 7.89 & 44.02 & 23.23 & -2.04 \\
129 & \text{NGC931} & \text{Mrk1040} & \text{Sy1} & 0.01631 & 70.7 & 7.41 & 44.41 & 21.09 & -1.18 \\
133 & \text{NGC973} & \text{UGC2048} & \text{Sy2} & 0.01566 & 67.9 & 8.48 & 44.08 & 22.49 & -2.58 \\
310 & \text{UGC3374} & \text{MCG+08-11-011} & \text{Sy1} & 0.02015 & 87.7 & 6.61 & 44.93 & 20.46 & 0.15 \\
382 & \text{Mrk79} & \text{UGC3973} & \text{Sy1} & 0.02211 & 96.3 & 7.61 & 44.54 & 20.00 & -1.25 \\
385 & \text{UGC3995B} & \text{MCG+5-19-1} & \text{Sy2} & 0.01590 & 69.0 & 7.97 & 43.92 & 23.92 & -2.22 \\
400 & \text{IC486} & \text{UGC4155} & \text{Sy1.9} & 0.02655 & 116.0 & 8.05 & 44.53 & 22.06 & -1.70 \\
451 & \text{IC2461} & \text{UGC4943} & \text{Sy2} & 0.00754 & 58.9 & 7.20 & 43.76 & 22.78 & -1.62 \\
484 & \text{NGC3079} & \text{UGC5387} & \text{Sy2} & 0.00350 & 20.6 & 6.38 & 43.60 & 24.56 & -0.95 \\
654 & \text{NGC4939} & \text{MCG-2-33-104} & \text{Sy2} & 0.01054 & 42.1 & 7.75 & 43.61 & 23.29 & -2.31 \\
665 & \text{NGC5033} & \text{UGC8307} & \text{Sy1.9} & 0.00276 & 19.1 & 7.75 & 42.22 & 20.00 & -3.70 \\
687 & \text{Z102-048} & \text{CGCG102-048} & \text{Sy2} & 0.02696 & 117.9 & 8.23 & 44.46 & 23.60 & -1.95 \\
733 & \text{NGC5674} & \text{UGC9369} & \text{Sy1.9} & 0.02477 & 108.1 & 7.81 & 44.21 & 22.84 & -1.77 \\
766 & \text{NGC5899} & \text{UGC9789} & \text{Sy2} & 0.00860 & 45.1 & 7.96 & 43.56 & 23.03 & -2.58 \\
828 & \text{NGC6232} & \text{UGC10537} & \text{Sy2} & 0.01489 & 64.5 & 7.35 & 44.10 & 24.35 & -1.43 \\
1162 & \text{UGC12138} & \text{MCG+01-57-016} & \text{Sy1} & 0.02494 & 108.8 & 7.08 & 44.21 & 20.00 & -1.04 \\
1184 & \text{NGC7479} & \text{UGC12343} & \text{Sy2} & 0.00710 & 36.8 & 7.58 & 43.41 & 24.16 & -2.35 \\
1198 & \text{NGC7682} & \text{UGC12622} & \text{Sy2} & 0.01706 & 74.0 & 7.84 & 44.41 & 24.27 & -1.61 \\
1202 & \text{UGC12741} & \text{CGCG497-048} & \text{Sy2} & 0.01799 & 78.1 & 7.62 & 44.19 & 23.82 & -1.61 \\
\hline
\end{tabular}
\tablefoot{
The BAT ID, AGN Type, redshift, distance, black hole mass, bolometric luminosity, hydrogen column density, and Eddington ratios are cited from the BASS \citep{Ricci2017, koss2022cat}. 
The most common catalog ID and alternative names are shown in Columns 2 and 3.
}
\end{table*}

\begin{figure}[!t]
\centering
\includegraphics[width=9cm]{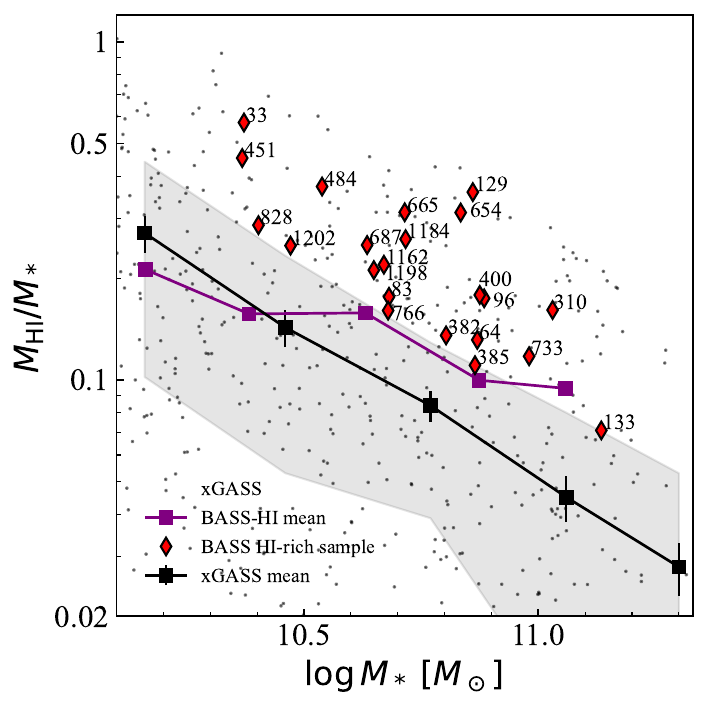}
\caption{\HI\ gas mass fraction of the sample as a function of stellar mass. 
The red diamonds show our sample with BAT IDs, and the purple squares show the binned mean gas fraction of the entire BASS-\HI-detected sample from single-dish measurements.
The black dot points are the xGASS sample \citep{Catinella2018}, and the black squares and the grey shaded regions represent the binned mean value and the 5$\sigma$ scatter of the same sample, respectively. 
}
\label{figure_mhi_mstar}
\centering
\end{figure}

Our sample was selected from the {\it Swift}/BAT AGN Spectroscopic Survey \citep[BASS\footnote{\url{https://www.bass-survey.com/}};][]{Koss2017, Ricci2017, Oh2018ApJS, koss2022overview}. BASS is a hard X-ray (14-195 keV) all-sky survey that includes the brightest X-ray AGNs, followed up by optical spectroscopy via ground telescopes. The hard X-ray emission at $E>10$~keV is less affected by star formation contamination as well as dust and gas obscuration, providing a relatively unbiased AGN sample in the local universe. The follow-up observations provide various measurements of the central SMBH properties, making BASS-selected AGNs ideal laboratories for probing the link between the host galaxy and AGNs.

We begin with the BASS DR2 parent sample \citep{koss2022cat}, restricting to nearby ($D < 120$~Mpc or $z<0.0275$)
galaxies to optimize for both angular resolution and sensitivity using the C-configuration of the Karl G. Jansky Very Large Array (VLA)\footnote{\url{https://public.nrao.edu/telescopes/vla/}}, which leaves 259 targets among the BASS DR2 sample of 858.
We chose galaxies that were previously detected via single-dish observations to ensure the targets have sufficient \HI\ gas to be resolved with reasonable observing time.
For this, we selected targets with reliable single-dish measurements based on the Extragalactic Distance Database \citep[EDD,\footnote{\url{https://edd.ifa.hawaii.edu}}][]{Courtois2009}, the Arecibo Legacy Fast ALFA Survey \citep[ALFALFA,\footnote{\url{https://egg.astro.cornell.edu/index.php/}}][]{Haynes2018}, and \citet{Springob2005}, making the BASS-\HI\ sample for which we have been collecting the spatially resolved \HI\ data.
These $\sim$100 targets are not biased toward any particular X-ray luminosity or stellar mass bins within the distance-limited BASS sample ($D < 120$~Mpc), and are quite uniformly distributed with ${L}_{\rm bol}\approx 10^{41.28}-10^{44.95}$~erg~s$^{-1}$, ${M}_{\rm BH}\approx 10^{5.45}-10^{9.05}M_\odot$ and ${M}_*\approx 10^{7.87}-10^{11.18}M_\odot$.
A more complete census of the BASS-\HI\ sample will be presented and discussed in greater detail in our forthcoming catalog paper (Kim et al., in prep.).

For this study, in which we particularly aim to specifically target the cases undergoing gas accretion, we selected 22 galaxies that are \HI-richer for a given stellar mass by 5$\sigma$ than the xGASS sample \citep[][]{Catinella2018}, a stellar-mass selected \HI\ single-dish survey (see Figure \ref{figure_mhi_mstar}).
These 22 targets have \HI\ masses, for a given stellar mass, comparable to those of \HI-rich xGASS galaxies, and such \HI-massive systems are found at similar frequencies in both samples.

The general properties of the sample, including AGN properties adopted from BASS DR2 such as BH masses, bolometric luminosities, and Eddington ratios \citep{koss2022cat}, are summarized in Table \ref{tab:bass-hi-rich-sample-bass}.
The BH masses were measured either from the velocity dispersion or broad lines \citep[see][for further details]{Mejia2022_MBH, Koss2022_MBH}.
The bolometric luminosity was calculated from the hard X-ray luminosity (14–150 keV) by applying a correction factor \citep[see][for further details]{koss2022overview}.
We also adopted the BASS stellar masses, which have been estimated by the Spectral Energy Distribution (SED) decomposition fitting. 
Details are described in \cite{Powell2018} \citep[see also][]{bar2019}.
The intrinsic X-ray luminosity of our sample obtained from \cite{Ricci2017}, ${L}_{\rm 2-10\,keV}$, ranges from $10^{40.89} - 10^{43.79}$ erg~s$^{-1}$, classifying our sample as low- to moderate-luminosity X-ray-selected AGNs.\footnote{The classification is adopted from \cite{Mountrichas2012}.}
Our sample encompasses a range of distances in the local universe ($20 \lesssim {D} \text{(Mpc)} <120$), hydrogen column density measured from X-ray spectra ($10^{20.0} < N_{H} \text{(cm}^{-2}\text{)} < 10^{24.56}$), and bolometric luminosities ($10^{42.22} < {L}_{\rm bol} \text{(erg/s)} < 10^{44.73}$), thereby representing \HI-rich Seyfert galaxies in the local universe.

\vspace{5mm}

\subsection{Observations}

To spatially resolve \HI\ gas with a sufficient number of synthesized beams, we conducted VLA C-configuration observations (VLA/20A-123, PI: A. Chung). To detect diffuse \HI\ gas within a reasonable amount of observational time (e.g., on-source time of $\sim1$hr), we aimed to reach a 1$\sigma$ \HI\ column density sensitivity of ${N}_{\rm HI} \approx 2 \times 10^{19} \text{cm}^{-2}$ over a 10 $\text{km}\;\text{s}^{-1}$ channel per synthesized beam of a few tens of arcseconds, offering an optimal balance between spatial resolution and sensitivity. More recently, to better probe the \HI\ gas in the target and its surroundings with deeper sensitivity, we also conducted VLA D-configuration observations (VLA/23B-079, PI: J. Kim), achieving a 3$\sigma$ \HI\ column density sensitivity of ${N}_{\rm HI} \approx 5 \times 10^{18} \text{cm}^{-2}$ over a 10 $\text{km}\;\text{s}^{-1}$ channel per beam. This deeper sensitivity allows us to detect more diffuse \HI\ gas around the targets. Additionally, we used archival data of sufficient sensitivity (e.g., $N_{\rm HI}\approx 10^{19-20}\text{cm}^{-2}$) and resolution (e.g., $\approx$20"), primarily utilizing the VLA data. 
Finally, we added two galaxies from other facilities: one case from the \HI-rich BASS sample observed with the Australia Telescope Compact Array (ATCA) as part of our southern hemisphere survey (ATCA/C3311, PI: O. I. Wong), and one case found in the archive of the Giant Metrewave Radio Telescope (GMRT). For both cases, the visibilities were reprocessed as the VLA data and included in our analysis.

In addition to our own VLA D-configuration data, we incorporated archival VLA D observations where available, aiming to investigate the surrounding gas environment while maintaining sufficient spatial resolution for detailed analysis of each target.
The observational details are given in Table \ref{tab:observational-properties}.

\subsection{Data reduction}

We used Common Astronomy Software Application \footnote{\url{https://casa.nrao.edu/}} \citep[\texttt{CASA v.6.5.4};][]{2007mcmulin, TheCASATeam_2022_nov} to reduce our spectral line observation. We utilized the VLA pipeline to calibrate the data for VLA observational projects 20A-123 and 23B-079, while manually calibrating the archive data observed before 2010 and the GMRT data.
For the ATCA data, we used the 
\texttt{Miriad v.4.3.8}\footnote{\url{https://www.atnf.csiro.au/computing/software/miriad/}} \citep[][]{miriad_sault_1995ASPC} for the calibration and the \texttt{CASA} for imaging.

The general calibration process was followed to apply phase correction, antenna delays, and band-pass corrections using the following tasks: \texttt{setjy}, \texttt{gencal}, \texttt{bandpass}, \texttt{gaincal}, \texttt{fluxscale}, and \texttt{applycal}.
Since the \HI\ spectral line resides in the L-band (1-2 GHz), the observations are affected by significant radio frequency interference (RFI). We removed RFI using \texttt{flagdata}, \texttt{tfcrop}, and \texttt{rflag}.

After calibration, the continuum was subtracted using the task \texttt{uvcontsub}, with first-order polynomial fitting from the line-free channels. Then, we imaged our data to spectral line cubes using CASA task \texttt{tclean}, with various weighting schemes. In most cases, natural weighting was used to recover the outermost diffuse gas distributions fully. Finally, for the coherence, all data cubes were smoothed to a width of $\sim$20 $\text{km}\;\text{s}^{-1}$, providing a good compromise between the signal-to-noise ratio and the study of kinematics, except for one case for which the initial channel resolution was adopted. NGC~262, which is nearly a face-on system with an \HI\ line width less than 100~km~s$^{-1}$, was imaged to a cube at a 10 $\text{km}\;\text{s}^{-1}$ to ensure sufficient spectral resolution. 

The mean 1$\sigma$ \HI\ column density sensitivities are 
$N_{\rm HI}=1.12\pm 0.62$ and $1.74\pm 2.23\times 10^{19}\textrm{cm}^{-2}$, for the VLA C and D combined final cube in our observations and archival data respectively, and the median column density sensitivity of the entire set is $N_{\rm HI}=8.05\times 10^{18}\textrm{cm}^{-2}$.

\begin{table*}
\caption{Overview of the observations.\label{tab:observational-properties}}
\centering
\begin{tabular}{ccccccc}
\hline\hline
BAT ID & Telescope & Project Code & Obs. Year/Month & $\sigma_{\rm cube}$ & Chan. Width & Synthesized Beam (P.A.) \\
& & & & ($\text{mJy}\;\text{beam}^{-1}$) & ($\text{km}\;\text{s}^{-1}$) & (arcsec × arcsec) (deg) \\
\hline
33 & VLA\tablefootmark{*} & AH417(C), AH372(D) & Nov90, Nov89 & 0.28 & 10.3 & 49.5 × 44.3 (+46.8) \\
64 & VLA\tablefootmark{*} & 20A-123\tablefootmark{a}, 23B-079\tablefootmark{b} & Mar20, Nov23 & 0.52 & 20.0 & 40.7 × 36.1 (+81.8) \\
83 & ATCA & C3311\tablefootmark{c} & Aug19 & 1.14 & 20.0 & 221.7 × 84.1 (+1.2) \\
96 & VLA\tablefootmark{*} & 20A-123\tablefootmark{a}, 23B-079\tablefootmark{b} & Mar20, Oct23 & 1.43 & 20.0 & 49.8 × 33.7 (+16.0) \\
129 & VLA\tablefootmark{*} & AL516(D) & Oct81, Jul00 & 0.60 & 20.6 & 59.6 × 52.8 (+72.9) \\
133 & GMRT & 20\_034 & Aug11 & 0.81 & 27.5 & 27.5 × 17.3 (+23.6) \\
310 & VLA & 23B-079\tablefootmark{b} & Nov23 & 0.64 & 20.0 & 71.7 × 55.7 (-66.0) \\
382 & VLA\tablefootmark{*} & 20A-123\tablefootmark{a}, 23B-079\tablefootmark{b} & Feb20, Nov23 & 0.46 & 20.0 & 38.5 × 36.0 (+84.6) \\
385 & VLA & AL516(D) & Aug00 & 1.23 & 20.6 & 102.8 × 50.5 (-68.1) \\
400 & VLA\tablefootmark{*} & 20A-123\tablefootmark{a}, 23B-079\tablefootmark{b} & Feb20, Oct23 & 0.45 & 20.0 & 32.5 × 22.5 (-65.5) \\
451 & VLA & AG645(D) & Feb03 & 0.91 & 20.6 & 65.1 × 56.5 (-82.1) \\
484 & WSRT & \cite{n3079_WSRT_2015MNRAS} & Jan04 & 0.12 & 16.5 & 32.9 × 29.1 (+4.3) \\
654 & VLA & 16A-269(C) & Mar16 & 0.41 & 20.0 & 22.5 × 15.4 (+6.5) \\
665 & VLA\tablefootmark{*} & AW701(C), AP270(D) & Aug93, Dec93 & 0.42 & 20.6 & 33.4 × 31.3 (+70.8) \\
687 & VLA\tablefootmark{*} & 20A-123\tablefootmark{a}, 23B-079\tablefootmark{b} & Apr20, Nov23 & 0.46 & 20.0 & 22.4 × 17.7 (-35.3) \\
733 & VLA\tablefootmark{*} & 20A-123\tablefootmark{a}, 23B-079\tablefootmark{b} & May20, Nov23 & 0.95 & 20.0 & 44.2 × 32.7 (+36.5) \\
766 & VLA & 23B-079\tablefootmark{b} & Nov23 & 0.60 & 20.0 & 53.5 × 50.0 (-26.3) \\
828 & VLA\tablefootmark{*} & 20A-123\tablefootmark{a}, 23B-079\tablefootmark{b} & May20, Oct23 & 0.48 & 20.0 & 43.0 × 27.4 (-38.2) \\
1162 & VLA\tablefootmark{*} & 20A-123\tablefootmark{a}, 23B-079\tablefootmark{b} & Mar20, Nov23 & 0.47 & 20.0 & 42.3 × 29.5 (+48.1) \\
1184 & VLA\tablefootmark{*} & 14B-396(C), AE175(D) & Nov14, Jun10 & 0.39 & 20.0 & 39.9 × 30.7 (+5.5) \\
1198 & VLA\tablefootmark{*} & AL542(C), AL516(D) & Sep01, Jul00 & 0.37 & 20.6 & 35.1 × 32.1 (-1.4) \\
1202 & VLA\tablefootmark{*} & 20A-123\tablefootmark{a}, 23B-079\tablefootmark{b} & Mar20, Jan24 & 0.50 & 20.0 & 35.0 × 32.4 (-80.1) \\
\hline
\end{tabular}
\tablefoot{The observational information of our sample for the VLA and ATCA observations, and collected archival data, including the GMRT. 
The channel width is smoothed to $\sim$20 km~s$^{-1}$, except NGC~262 and NGC~3079 for which the initial channel resolutions were used. Our own observations are footnoted - \tablefoottext{a}{PI: Aeree Chung}, \tablefoottext{b}{PI: Jeein Kim}, \tablefoottext{c}{PI: O. Ivy Wong}, \tablefoottext{*}{C- and D-configuration data combined.}
}
\end{table*}

\vspace{5mm}

\section{Results} \label{sec: results}

\subsection{\HI\ morphologies and kinematics} \label{subsec: morph}

\begin{figure*}[pth]
    \centering
    \includegraphics[width=12.8cm]{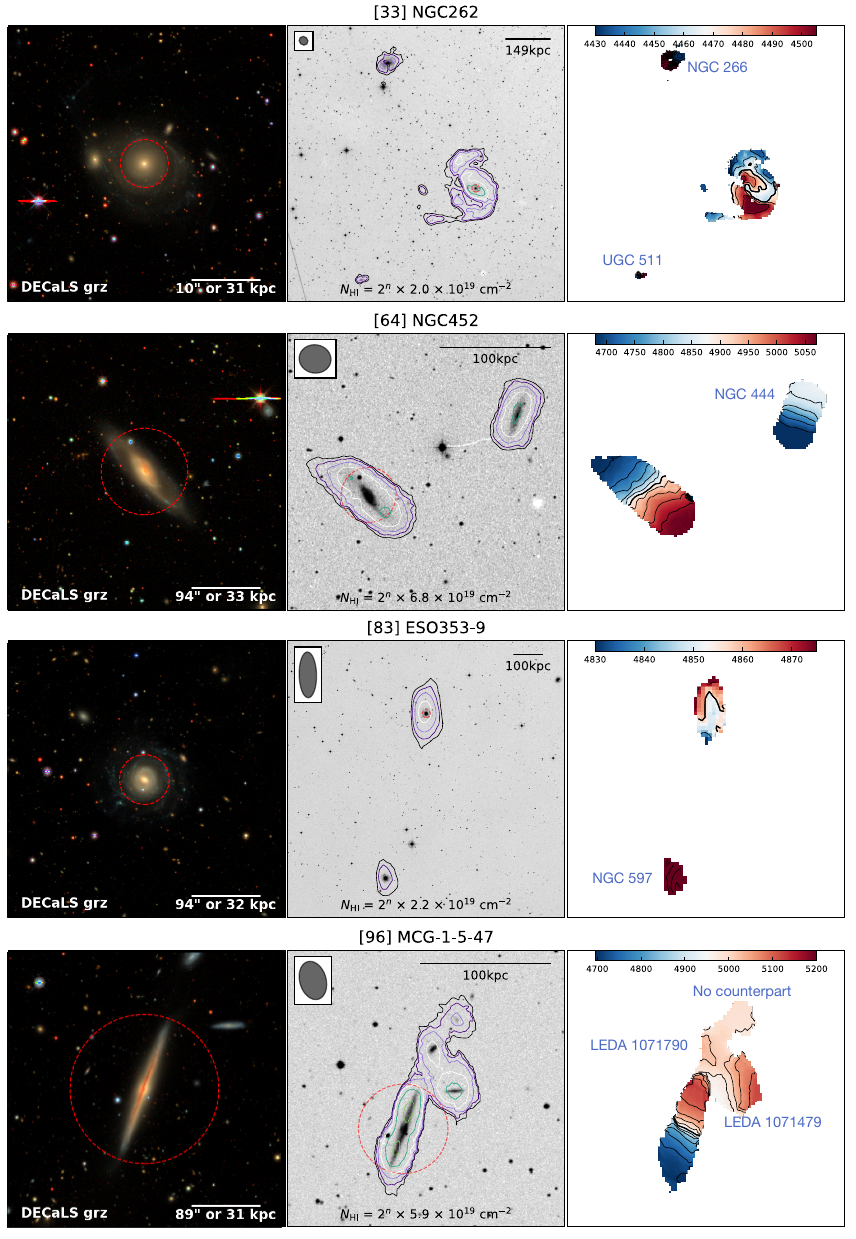}
    \caption{
    Images of the first four galaxies. The rest of the samples are presented in the appendix \ref{appendix:continued_mom_maps}.
    ({Left}) The optical color image. The angular and physical scale bar is shown on the bottom right of each panel. (Center) The integrated \HI\ intensity contours overlaid on the DSS-1 image.
    The synthesized beam is shown on the upper left of each panel.
    The contour levels are (the lowest contour level in $\text{cm}^{-2}$  which is presented at the bottom)$\times2^{n}$, where n=0, 1, 2, 3, and so on. Various contour colors are used to reveal the inner stellar structure clearly, and they are not associated with specific column densities. 
    The scale bar shows 50, 100, or 150 kpc at the given distances of each galaxy.
    ({Right}) The moment 1 (velocity field) map.
    The central velocity is listed at the bottom and indicated by a thick line. 
    In the approaching/receding side (blue/red) of the disk, velocities are indicated with solid/dashed lines, with a separation of 40 $\text{km}\;\text{s}^{-1}$, except for NGC~262, ESO353-G009, and NGC~6232, for which are separated by 15 $\text{km}\;\text{s}^{-1}$.
    The colorbar on the top of the panel is in the unit of km~s$^{-1}$. The names of \HI-detected neighbor(s) are noted in blue.
    }\label{fig_mom_C_A}
    \centering
\end{figure*}

\begin{figure}[hbt]
    \centering
    \includegraphics[width=9cm]{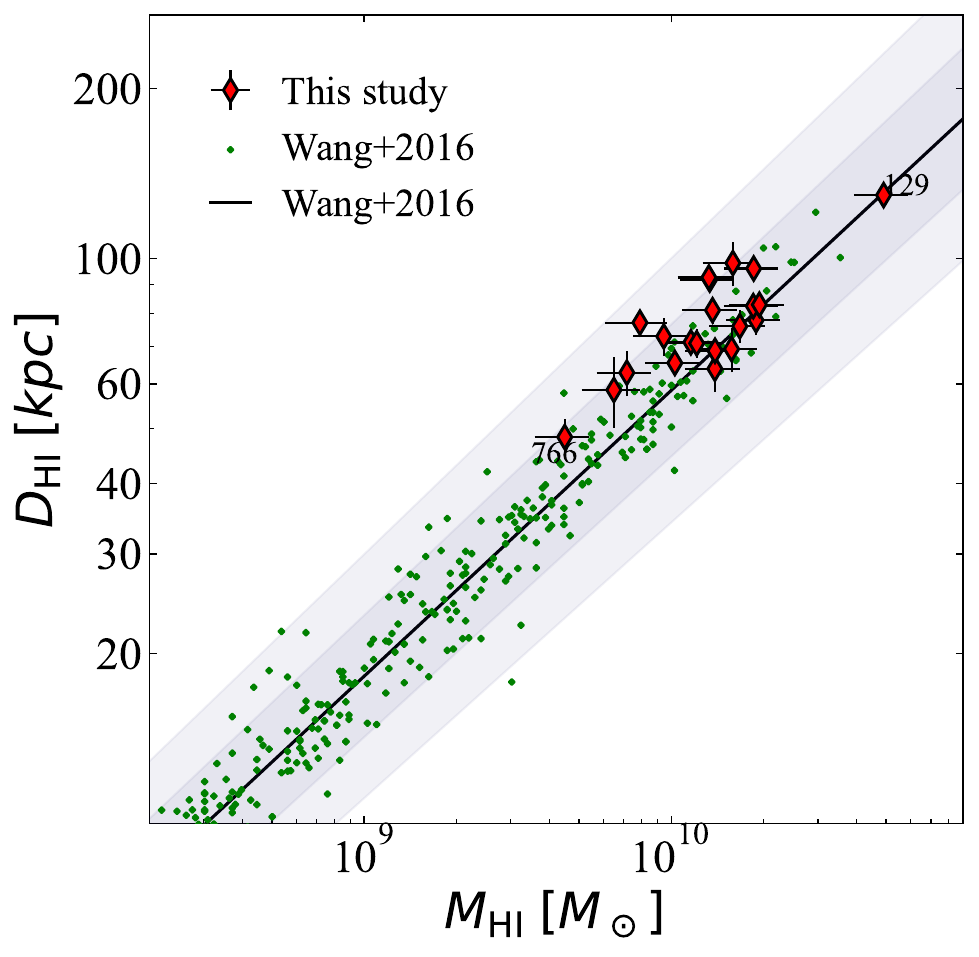}
    \caption{\HI\ mass and size scaling relation. The green dots represent $\sim$500 nearby galaxies analyzed by \cite{Wang2016}. The solid black line shows the best fit of \cite{Wang2016}’s sample, and grey colors show its 3$\sigma$ and 6$\sigma$ scatters. The errorbars of ${D}_{\rm HI}$ and ${M}_{\rm HI}$ measurements are plotted. The red diamonds represent the sample for this study. Most of our targets follow the scaling relation within a 3$\sigma$ scatter and are clustered in the high ${M}_{\rm HI}$–${D}_{\rm HI}$ regime. The objects that deviate most significantly from the rest of our sample are labeled with their BAT~IDs.
    }
    \label{figure_size_mass}
    \centering
\end{figure}

Integrated intensity and velocity maps were produced by taking the 0th and the intensity-weighted 1st moments of the cube, respectively, along the velocity axis.
The moment maps were produced by taking moments of the data cubes along the frequency axis using the Astronomical Image Processing System\footnote{\url{https://www.aips.nrao.edu}} \citep[\texttt{AIPS};][]{aips_1990, aips_Greisen_2003} task \texttt{MOMNT}. During mask creation in \texttt{MOMNT}, we applied Gaussian smoothing in the spatial direction and Hanning smoothing in the frequency direction to maximize the signal-to-noise ratio. We typically adopted a cutoff value of 1.5–2.5 times the rms.

The results of four galaxies are presented in Figure \ref{fig_mom_C_A} and the rest of the sample are presented in Fig.\ref{fig_mom_C_B} in the appendix \ref{appendix:continued_mom_maps}. On the left of each panel, we show color images mainly from the Dark Energy Camera Legacy Survey (DECaLS), using the stretch settings described in \cite{BASS_Miguel_Parra_morph_2025}. When the DECaLS images are not available, we show the Panoramic Survey Telescope \& Rapid Response System (Pan-STARRS)\footnote{\url{https://www2.ifa.hawaii.edu/research/Pan-STARRS.shtml}} colored images. The optical reference of each panel is shown on the bottom left.
The center of each panel presents \HI\ moment-0 contours overlaid on r-band images from the Digitized Sky Survey (DSS-1)\footnote{\url{https://irsa.ipac.caltech.edu/data/DSS/overview.html}}. On the right, the \HI\ moment-1 (velocity field) maps are displayed. Most of the galaxies are shown with a larger field of view compared to the optical image, to show their surrounding environments.

When available, we combined the VLA C-and D-configuration to optimize the balance between spatial resolution and sensitivity, resulting in a synthesized beam of approximately $\sim30-40$". For the ATCA data, we used only the inner five array elements to avoid excessively elongated beam shapes, yielding a beam size of 221.70"$\times$84.13". For the GMRT data, we smoothed the beam to $\sim20$" to maximize flux recovery. 

Our sample is generally located in a group-like gas-rich environment, as shown in Figure \ref{fig_mom_C_A} and Fig.\ref{fig_mom_C_B} (discussed in more detail in Section \ref{discuss:local_env}). More than half of the sample ($\sim$80\%; 18 out of 22) have gas-rich companions, mainly with an optical counterpart.

\vspace{5mm}

\subsection{Measured \HI\ properties of the sample} \label{sec: measurements}

The \HI\ flux is measured as below:
\begin{equation}
S_{\rm HI}\;[\text{Jy}\;\text{km}\;\text{s}^{-1}] = \Sigma {F}_{\rm HI,\textit{i}} \times \Delta {V},
\end{equation}
where ${F}_{\rm HI}$ is the integrated flux density of one channel in $\text{Jy}$, and $ \Delta {V} $ is the velocity separation between each channel in $\text{km}\;\text{s}^{-1}$.
The uncertainty in the total flux is estimated as follows:
\begin{equation}
\sigma_{S_{\rm HI}}\;[\text{Jy}\;\text{km}\;\text{s}^{-1}] = 
\sqrt{\Sigma (\sigma_{\rm HI, \textit{i}})^2}
\times \Delta {V},
\end{equation}
where $\sigma_{\rm HI, \textit{i}}$ is the RMS in each channel in units of $\text{Jy}$, accounting for the number of beams included in the flux measurement ($=\sigma_{\rm cube}\times \sqrt{N_{\rm beam}}$).
The \HI\ mass is calculated as below \citep{Horellou2001}:
\begin{equation}
\label{eq_HI_mass_Horellou01}
{M}_{\rm HI}\;[{M}_\odot] = 2.356\times 10^5\;{D}^2_{\rm Mpc}\; S_{\rm HI},
\end{equation}
where ${D}_{\rm Mpc}$ is the distance in Mpc from Table \ref{tab:bass-hi-rich-sample-bass} and $S_{\rm HI}$ is the integrated \HI\ flux in Jy~km~s$^{-1}$.
The uncertainty in \HI\ mass is measured as below:
\begin{equation}
\sigma_{{M}_{\rm HI}}\;[{M}_\odot] = {M}_{\rm HI} \times \sqrt{2\times (\frac{\sigma_{{D}_{\rm Mpc}}}{{D}_{\rm Mpc}})^2 + (\frac{\sigma_{S_{\rm HI}}}{S_{\rm HI}})^2},
\end{equation}
For the distance uncertainty $\sigma_{{D}_{\rm Mpc}}$, we adopt the mode value from the extragalactic distance survey  \citep[Cosmic Flow-3;][]{Tully2016AJ_CosmicFlow3}, which is 20\% for the distance within 120~Mpc. 
The typical extragalactic distance uncertainty varies depending on the methodology (e.g., a few \% of calibrated the tip of the red giant branch (TRGB); \citeauthor{Freedman2019_TRGB} \citeyear{Freedman2019_TRGB}; up to $\sim$20\% for the Tully-Fisher relation; \citeauthor{Tully2016AJ_CosmicFlow3} \citeyear{Tully2016AJ_CosmicFlow3}), and we adopt the most conservative value in this work.

The linewidths, $w_{20}$ and $w_{50}$, are measured at 20\% and 50\% of the peak fluxes on each side of the global line profiles. We adopt the velocity resolution of each cube as the uncertainty for $w_{20}$, $w_{50}$, and $v_{\rm sys}$. The systemic \HI\ velocity ($v_{\rm sys}$) is determined by averaging the low and high velocity limits used to define $w_{20}$ and $w_{50}$.

\begin{table*}
\caption{Measured \HI\ properties \label{tab:measured_properties}}
\centering
\begin{tabular}{ccccccccc}
\hline\hline
BAT ID & $v_{\rm sys}$ & $w_{20}$  & $w_{50}$ & $D_{\rm HI}$  & $S_{\rm HI}$ & $\log {M}_{\rm HI}$ & Neighbors\tablefootmark{1} & HI char\tablefootmark{2} \\
 &  $ (km\; s^{-1})$ & $(km\; s^{-1})$ & $(km\; s^{-1})$ & (arcmin) & ($Jy\; km \; s^{-1}$) & $({M}_\odot)$  & & \\
\hline
33 & 4469.3 & 95.4 & 66.7 & 2.60 $\pm$ 0.06 & 19.51 $\pm$ 0.17 & 10.27$^{\scriptscriptstyle +0.08}_{\scriptscriptstyle -0.10}$ & 2/11 & T/C \\
64\tablefootmark{a} & 4886.8 & 491.8 & 455.0 & 1.90 $\pm$ 0.08 & 10.76 $\pm$ 0.29 & 10.13$^{\scriptscriptstyle +0.08}_{\scriptscriptstyle -0.10}$ & 1/1 & L  \\
83 & 4856.6 & 140.7 & 79.4 &   & 10.67 $\pm$ 0.37 & 10.10$^{\scriptscriptstyle +0.08}_{\scriptscriptstyle -0.10}$ & 1/14 &   \\
96 & 4935.5 & 599.7 & 521.5 & 1.84 $\pm$ 0.09 & 15.19 $\pm$ 0.31 & 10.28$^{\scriptscriptstyle +0.08}_{\scriptscriptstyle -0.10}$ & 3\tablefootmark{b}/7 & C \\
129 & 4910.7 & 466.1 & 424.5 & 3.15 $\pm$ 0.07 & 41.70 $\pm$ 0.37 & 10.69$^{\scriptscriptstyle +0.08}_{\scriptscriptstyle -0.10}$ & 5/28 & C \\
133 & 4801.3 & 608.1 & 578.5 & 1.95 $\pm$ 0.04 & 7.27 $\pm$ 0.79 & 9.90$^{\scriptscriptstyle +0.09}_{\scriptscriptstyle -0.11}$ & 1/24 &   \\
310 & 6008.1 & 321.1 & 285.0 & 1.93 $\pm$ 0.14 & 8.75 $\pm$ 0.30 & 10.20$^{\scriptscriptstyle +0.08}_{\scriptscriptstyle -0.10}$ & 1\tablefootmark{b}/- &   \\
382 & 6512.7 & 196.2 & 165.6 & 1.30 $\pm$ 0.12 & 4.32 $\pm$ 0.18 & 9.98$^{\scriptscriptstyle +0.08}_{\scriptscriptstyle -0.10}$ & 3\tablefootmark{b}/4 & L \\
385 & 4672.8 & 443.3 & 412.1 & 1.46 $\pm$ 0.22 & 5.79 $\pm$ 0.47 & 9.81$^{\scriptscriptstyle +0.08}_{\scriptscriptstyle -0.11}$ & 2/4 & C \\
400 & 7809.8 & 419.4 & 305.5 & 1.12 $\pm$ 0.10 & 5.27 $\pm$ 0.22 & 10.22$^{\scriptscriptstyle +0.08}_{\scriptscriptstyle -0.10}$ & 3/14 & T \\
451 & 2239.5 & 379.6 & 347.2 & 1.83 $\pm$ 0.14 & 8.76 $\pm$ 0.28 & 9.85$^{\scriptscriptstyle +0.08}_{\scriptscriptstyle -0.10}$ & -/2 & L \\
484\tablefootmark{a} & 1114.2 & 471.2 & 432.8 & 5.93 $\pm$ 0.02& 115.73 $\pm$ 0.21 & 10.06$^{\scriptscriptstyle +0.08}_{\scriptscriptstyle -0.10}$ & 7\tablefootmark{b}/14 & C \\
654 & 3079.1 & 455.7 & 437.2 & 3.75 $\pm$ 0.02 & 31.92 $\pm$ 0.81 & 10.12$^{\scriptscriptstyle +0.08}_{\scriptscriptstyle -0.10}$ & -/8 &   \\
665 & 880.4 & 456.2 & 423.9 & 7.42 $\pm$ 0.02 & 214.99 $\pm$ 1.20 & 10.27$^{\scriptscriptstyle +0.08}_{\scriptscriptstyle -0.10}$ & 4/41 &   \\
687 & 7960.5 & 538.6 & 91.2 & 1.35 $\pm$ 0.06 & 4.04 $\pm$ 0.25 & 10.12$^{\scriptscriptstyle +0.08}_{\scriptscriptstyle -0.10}$ & 1/5 & T/L \\
733 & 7299.4 & 271.2 & 236.9 & 1.10 $\pm$ 0.14 & 5.70 $\pm$ 0.34 & 10.20$^{\scriptscriptstyle +0.08}_{\scriptscriptstyle -0.10}$ & -/9 &   \\
766 & 2593.5 & 494.9 & 465.8 & 1.84 $\pm$ 0.12 & 9.38 $\pm$ 0.30 & 9.65$^{\scriptscriptstyle +0.08}_{\scriptscriptstyle -0.10}$ & 1/3 &   \\
828 & 4402.0 & 154.2 & 136.5 & 1.74 $\pm$ 0.08 & 10.47 $\pm$ 0.26 & 10.01$^{\scriptscriptstyle +0.08}_{\scriptscriptstyle -0.10}$ & 3/- &   \\
1162 & 7348.7 & 222.3 & 144.9 & 1.01 $\pm$ 0.14 & 4.97 $\pm$ 0.19 & 10.14$^{\scriptscriptstyle +0.08}_{\scriptscriptstyle -0.10}$ & 3\tablefootmark{b}/6 &   \\
1184 & 2368.0 & 378.4 & 348.3 & 3.21 $\pm$ 0.04 & 43.46 $\pm$ 0.48 & 10.14$^{\scriptscriptstyle +0.08}_{\scriptscriptstyle -0.10}$ & -/4 &   \\
1198 & 5032.7 & 236.7 & 201.3 & 1.93 $\pm$ 0.07 & 14.99 $\pm$ 0.38 & 10.29$^{\scriptscriptstyle +0.08}_{\scriptscriptstyle -0.10}$ & 3/14 & C \\
1202 & 5148.4 & 413.2 & 374.0 & 1.56 $\pm$ 0.09 & 8.41 $\pm$ 0.29 & 10.08$^{\scriptscriptstyle +0.08}_{\scriptscriptstyle -0.10}$ & 3\tablefootmark{b}/1 & C \\
\hline
\end{tabular}
\tablefoot{Velocities are provided in radio definitions. The $v_{\rm sys}$ represents the systematic velocity of the cube, which was measured by averaging $w_{20}$ and $w_{50}$. The $w_{20}$ and $w_{50}$ are the line widths that were measured at 20\% and 50\% of the peak fluxes on each side of the line profiles. 
The \HI\ size ${D}_{\rm HI}$ is measured at 
$\Sigma_{\rm HI}=1\;{M}_\odot \; {\rm pc}^{-2}$.
The integrated \HI\ flux $S_{\rm HI}$ were converted to \HI\ mass ${M}_{\rm HI}$ by equation \ref{eq_HI_mass_Horellou01}. The distances were adopted from the BASS DR2 survey \citep{koss2022cat}. \\
\tablefoottext{1}{(Left): The number of \HI-detected neighbors within a projected radius of $\sim$450~kpc and a velocity range of $\pm$300~km~s$^{-1}$ around each of our sample galaxies is shown. (Right): The number of neighboring objects identified via $\textrm{SIMBAD}$\citep{simbad2000} within a 1~Mpc radius and $\pm$500~km~s$^{-1}$} velocity range around each target galaxy is listed.
\tablefoottext{2}{\HI\ characteristics: ``T'' indicates the presence of one-sided \HI\ gas extent in moment maps; ``C'' denotes spatial and/or kinematical gas connections; ``L'' refers to the lopsidedness of \HI\ line profiles. See Section \ref{sec: measurements} for further details. }
\tablefoottext{a}{\HI\ absorption detected.} \tablefoottext{b}{A gas cloud with no optical counterpart was detected.} \tablefoottext{c}{See Table 2 in \cite{n3079_WSRT_2015MNRAS} for details.}
}
\end{table*}

We derive surface density profiles by fitting 3D tilted-ring models using the 3D-Based Analysis of Rotating Object via Line Observations \citep[\textrm{3D-Barolo}\footnote{\url{https://editeodoro.github.io/Bbarolo/}};][]{DiTeodoro2015}. We performed 3D-Barolo fitting on the full \HI\ cubes containing both signal and noise. To improve computational efficiency, we used mask cubes - containing the location of \HI\ emission identified by the 2nd version of the Source Finding Algorithm  \citep[$\textrm{SoFiA-2}$\footnote{\url{https://gitlab.com/SoFiA-Admin/SoFiA-2}};][]{Westmeier2021} - to guide the fitting process.
For each galaxy, the fitting was performed in several steps as follows. First, we applied \textrm{3D-Barolo} to the inner disk, where the rotation is fairly axisymmetric and regular. Initial guesses for the galactic center (XPOS, YPOS), position angle (PA), and inclination (INC) were taken from SoFiA, while the systemic velocity and rotational velocity (VSYS and VROT) were estimated from the \HI\ linewidths. The ring width (RADSEP) was set to approximately one quarter to one third of the synthesized beam, following the principle of Nyquist sampling but avoiding oversampling. In this step, XPOS, YPOS, and VSYS were left free so that their optimized values could be determined by the centroid of the fitting process. After fixing XPOS, YPOS, and VSYS to the values obtained in the first step, we then ran \textrm{3D-Barolo} on the entire disk, allowing VROT, PA, and INC to vary as a function of radius and yielding a corresponding density profile.

The \HI\ extent ${D}_{\rm HI}$ is measured at the radii where the azimuthally averaged \HI\ surface density $\Sigma_{\rm HI}$ drops to $1\;{M}_\odot \; {\rm pc}^{-2}$. If there is more than one radius that meets this criterion (e.g., in the case where an \HI\ hole or dip is present), the outermost radius is taken. The measurement uncertainty of ${D}_{\rm HI}$ represents the width of the ring for the ellipse fitting, which is approximately 1/4 to 1/3 of the synthesized beam size. This width is chosen to prevent undersampling following the Nyquist sampling theorem. BAT~ID 83, the only case observed using the ATCA among the sample of this work, was omitted from ${D}_{\rm HI}$ measurement since the map is only marginally resolved with $\sim$2-3 beams.

${D}_{\rm HI}$ measured this way often does not well represent the true extent of \HI\ in those cases where the \HI\ distribution is significantly asymmetric and/or there is more diffuse \HI\ gas with a large extent. Therefore we also define ${D}_{\rm HI, ext}$, the largest extent that can be measured in projection at ${N}_{\rm HI}=5\times10^{19}\text{cm}^{-2}$, which is deep enough to reveal tidally disturbed features and also ensure high signal-to-noise (SNR$\gtrsim3$) for C- and D-configuration combined data \citep[e.g.,][]{HI_Rogue_Gallery_Hibbard2001}. In Section \ref{sec:discuss_sec}, ${D}_{\rm HI, ext}$ will be further probed along with some AGN and star-forming properties of the host. We adopt the mean of the major and the minor synthesized beam as the uncertainty of ${D}_{\rm HI, ext}$.

The measured \HI\ properties are listed in Table \ref{tab:measured_properties}. To better characterize the environments, we also provide the number of neighboring galaxies and their \HI\ characteristics. The number of \HI-detected neighbors in each cube is defined as the galaxies located within a projected radius of $\sim$450~kpc and within $\pm$300~km~s$^{-1}$ of the systemic velocity of each target. In addition, we queried $\textrm{SIMBAD}$\footnote{\url{http://simbad.cds.unistra.fr/simbad/}} to identify optically known neighboring systems within 1~Mpc and within $\pm$500~km~s$^{-1}$. For three targets - BAT~IDs 33, 400, and 687 - we identified one-sided \HI\ extensions, denoted as ``T'' in Table \ref{tab:measured_properties}. These features appear as tail-like structures extending preferentially in one direction compared to the optical disk (Figs. \ref{fig_mom_C_A} and \ref{fig_mom_C_B}), and they show no connection to neighboring galaxies. We also denote as ``C'' the cases where a gas connection is present. These include systems exhibiting a common large-scale \HI\ envelope or an \HI\ bridge in the spatial distribution (Figs. \ref{fig_mom_C_A} and \ref{fig_mom_C_B}), or in velocity space (Fig. \ref{fig_PVD_appendix}).
Finally, galaxies with an asymmetry parameter $E$ ($=10(1-f)$, where $f$ is the flux ratio of the two horns of the \HI\ line profile; \citeauthor{Bournaud2005_lopsidedness} \citeyear{Bournaud2005_lopsidedness}), exceeding 2, were labeled as ``L'' in Table \ref{tab:measured_properties}. \cite{Bournaud2005_lopsidedness} find that two thirds of field spirals have $E\ge1$, whereas only one quarter show relatively high lopsidedness with $E\ge2$.

On average, we recover approximately 20\% more \HI\ emission compared to previous single-dish observations. This result is not surprising, given that our targets are relatively nearby and exhibit extended \HI\ distributions, often larger than the beam size of single-dish telescopes. In some cases, the single-dish measurements may have been contaminated by neighboring galaxies, leading to an overestimation of the \HI\ mass originally attributed to our targets. Nevertheless, even in such cases, our targets remain the dominant \HI\ sources in their environments, and their classification as \HI-rich still holds.

There are two cases (BAT~IDs 385 and 766) for which our interferometric observations did not fully recover the single-dish fluxes. For BAT~385, the array data may miss some emission, with discrepancies of up to $\sim$0.2~dex compared to the single-dish measurements. In the case of BAT~766, the total \HI\ flux within the GBT beam agrees with the VLA measurement; however, once spatially resolved, the galaxy itself appears to contain less \HI\ than inferred from the single-dish data, likely due to contamination from a nearby companion. Nevertheless, both galaxies still have at least the average \HI\ mass expected for their stellar masses when compared with the xGASS sample. More importantly, they remain the \HI-richest galaxies in their respective local environments, and we therefore include them in our analysis.

\subsection{\HI\ size-mass relation}

Figure \ref{figure_size_mass} shows the BASS-\HI-rich galaxies on the \HI\ size-mass (${D}_{\rm HI}-{M}_{\rm HI}$) scaling relation. This relation is one of the tightest scaling relations known, with a scatter $\sigma\sim0.06$ dex, which does not change much with the gas fraction or the luminosity of galaxies \citep{Wang2016}. As shown in Figure \ref{figure_size_mass}, the majority of our galaxies follow the scaling relation within 3$\sigma$ scatter.

That is, the global integrated \HI\ mass and flux of most of our sample are comparable to those of other normal (inactive) galaxies, with similar extents of gas distribution (see Figure \ref{figure_size_mass}). This is consistent with previous single-dish \HI\ studies of AGN samples. For example, \cite{Ho2008_ApJ} have shown that for Sb and later type spirals, the integrated \HI\ gas content is generally similar between Type~1 Seyfert galaxies and non-AGN galaxies, and there are no significant differences in their integrated \HI\ properties (e.g., the Tully-Fisher relation). This relationship appears to hold between the \HI\ size, measured at $1\;{M}_\odot \; {\rm pc}^{-2}$, and the total \HI\ mass for the sample of \HI-rich AGN hosts.

\subsection{Star-forming properties of the sample}

Star formation rates (SFRs) for BASS AGNs have been reported by \citeauthor{Ichikawa2019} (\citeyear{Ichikawa2019}; see also \citeauthor{Shimizu2015} \citeyear{Shimizu2015}; \citeauthor{Ichikawa2017ApJ} \citeyear{Ichikawa2017ApJ}), who used infrared SED modeling to distinguish the AGN and star formation contribution of the galaxy. The SFR of individual galaxies has been measured from the decomposed star-forming component. Since the SED fitting is dominated by the host-galaxy template, the uncertainty in the SFR is primarily driven by the uncertainty in the FIR flux density. Therefore, we adopt the measurement errors in the FIR flux densities at 70, 90, and 160 $\mu$m as the uncertainties on the SFR. For targets for which the FIR-flux measurement errors are not available, we conservatively assume a 10\% error, slightly larger than the typical uncertainties. For BAT~33 and BAT~64, which were not included in the study of \cite{Ichikawa2019}, we adopted SFR values from \cite{Leroy2019ApJS_SFR}.

\begin{figure}[!t]
    \centering
    \includegraphics[width=9.1cm]{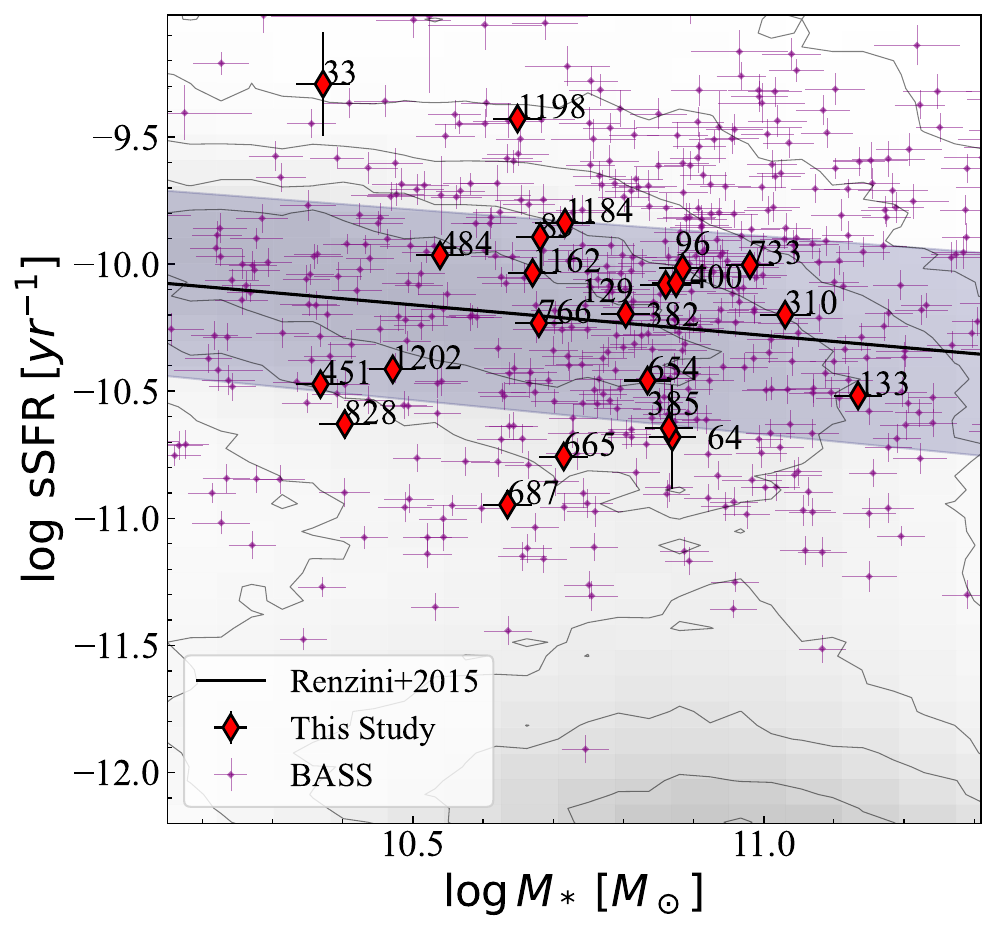}
    \caption{Global star formation properties of the sample. The black solid line shows the scaling relation from \cite{Renzini2015_sfms} with 3$\sigma$ scatter (shaded area). The background data (greyscale) is from the MPA-JHU catalog \citep{Brinchmann2004}, where the contours represent 2D histograms corresponding to approximately 30, 100, 300, 500, 700, 800, and 841 thousand galaxies. The purple points represent all the BASS galaxies in which the star formation rate was measured \citep{Ichikawa2019}. The red diamonds represent our sample, with BAT IDs.}
    \label{figure_sfms}
    \centering
\end{figure}

Figure \ref{figure_sfms} shows the relation between the specific SFR (sSFR=SFR$/M_\odot$) and stellar mass for our sample overlaid on the MPA-JHU catalog \citep{Brinchmann2004}. The background contours represent 2D histograms of MPA-JHU galaxies, which correspond to $\sim$30, 100, 300, 500, 700, 800, and 841 thousand galaxies, respectively. The solid black line indicates the star-forming main sequence modeled by \cite{Renzini2015_sfms} and the shaded area represents a 3$\sigma$ scatter. The red diamonds represent our galaxies. A few outliers are clearly noticeable. Two galaxies, BAT~33 (NGC~262) and BAT~1198 (NGC~7682), deviate from the scaling relation toward higher sSFR, whereas BAT~687 (Z~102-048) exhibits a lower sSFR than expected for its stellar mass. The latter case shows a stretched, one-sided \HI\ feature with no optical counterpart, which may suggest recent gas accretion possibly related to tidal interactions. Such gas exchange with neighboring systems could have perturbed the star-forming gas, potentially contributing to its low star-formation efficiency. Meanwhile, BAT~33 and BAT~1198 may have already undergone tidal interactions, given their proximity to neighboring galaxies, which could have triggered enhanced star formation. The remaining galaxies are located along the star-forming main sequence, within the expected scatter. This indicates that our galaxies are globally star-forming and not in the phase of quenching.

\vspace{5mm}

\section{Discussion} \label{sec:discuss_sec}

\subsection{Cool gas supply of the AGN hosts}  \label{discuss:local_env}

\begin{figure*}[th]
    \centering
    \includegraphics[width=\textwidth]{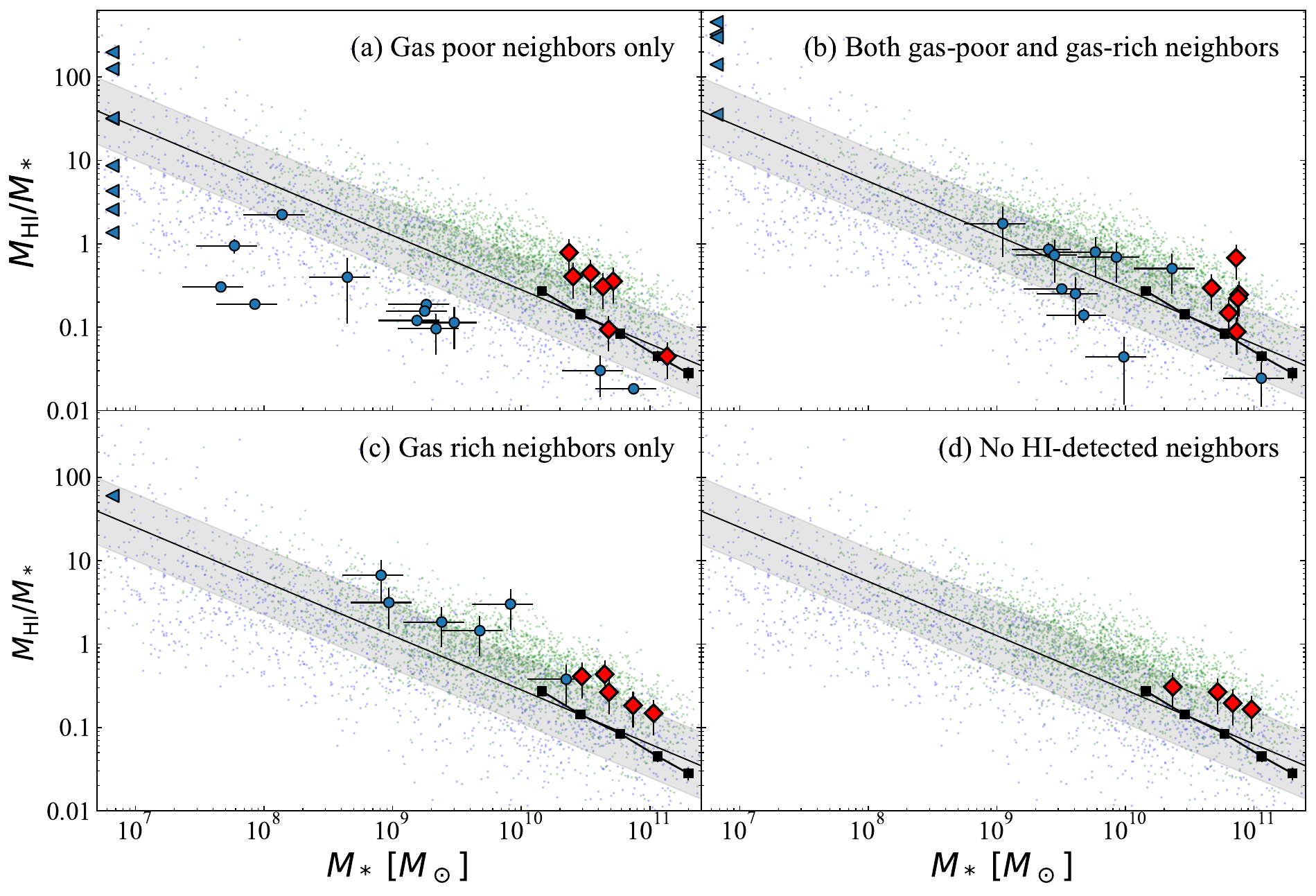}
    \caption{\HI-to-stellar mass fraction vs. stellar mass for our targets (red diamonds) and their neighbors (blue circles). 
    Panel (a) shows a subsample in which all \HI-detected neighbors are gas poor for their given stellar mass. Panel (b) shows galaxies in which both gas-poor and gas-rich neighbors are observed. 
    Panel (c) shows the cases with only gas-rich neighbors.
    Panel (d) shows the hosts with no other neighbors with \HI\ detected in their surroundings. 
    \HI\ clouds detected in our observations but not visible in the 2MASS $\text{K}_s$-band are shown as triangles.
    The upper limit on their stellar mass was estimated by converting the 2MASS $\text{K}_s$-band surface brightness limit using the scaling relation from \cite{Wen2013}, yielding a limit of $10^{6.82}{M}_\odot$.
    The solid black line represents the scaling relation from the \HI-selected spiral sample in the \HI\ Parkes All-Sky Survey Catalog \citep[HICAT;][]{Meyer2004, Parkash2018ApJ}, with the gray shaded area indicating the $1\sigma$ scatter. Background green points show HICAT data; blue points represent Nançay data \citep{vanDriel2016}; and black squares indicate binned mean values from the xGASS sample \citep{Catinella2018}, as in Figure \ref{figure_mhi_mstar}. 
    Except for group (c), other host galaxies are likely to have acquired gas from their neighbors. }
    \label{figure_gas_fraction_neighbors}
    \centering
\end{figure*}

Extended gas tails and bridges can be indications of ongoing or past tidal interactions. As mentioned, 12 galaxies among our sample show lopsidedness, one-sided tails, or connected gas bridges, likely due to tidal interactions with their companions. Six galaxies (BAT~IDs~33, 96, 129, 385, 1198, and 1202) are the most evident cases as seen by the common gas envelope with their neighbors. The high frequency of tidal interactions in our Seyfert sample is consistent with previous \HI\ studies. For instance, \citeauthor{Kuo2008} (\citeyear{Kuo2008}, see also \citeauthor{Haan2008} \citeyear{Haan2008}) found that a significant fraction ($\sim$94\%) of local Seyfert galaxies show signs of tidal disturbances when compared with a control sample of inactive galaxies \citep{Tang2008}, among which only 3 out of 20 ($\sim$15\%) exhibit disturbed \HI\ morphology.

In addition, 18 out of 22 sample galaxies ($\sim$80\%) have neighbors that are also detected in \HI. In comparison, the xGASS group catalog \citep{Yang2007, Catinella2018} for stellar masses matched galaxies to our sample ($10^{10.37} < {M}_*/M_{\odot} < 10^{11.14}$) finds $\sim$67\% to be in the group environment, among which $\sim$31\% is in the group center. Considering that not all the members of xGASS groups are detected in \HI, our selected subsample shows a high probability of finding neighbors in the same stellar mass range.

Intriguingly, our targets are the most massive systems in their respective groups in the majority of our sample. This makes our AGN sample likely the host of gas accretion if there were any matter exchange among the members.
In Figure \ref{figure_gas_fraction_neighbors}, our AGN hosts and their neighbors are shown in relation to the atomic gas fraction and stellar mass of galaxies. 
If our galaxies have acquired gas from their neighbors, then at least one of the neighboring systems would be expected to contain less \HI\ than predicted from its stellar mass, or even to fall below the \HI\ detection limit. Accordingly, we divide Figure \ref{figure_gas_fraction_neighbors} into four panels: (a) systems in which all neighbors lie below the expected \HI\ gas-fraction relation; (b) systems with a mix of gas-poor and gas-rich neighbors; (c) systems in which all neighbors are gas-rich; and (d) systems with no \HI-detected neighbors. Even in case (d), each system still has at least one optical neighbor, suggesting that these companions likely contain less \HI\ than expected.

For neighboring galaxies, which lack stellar mass measurements, we derive their stellar masses coherently by converting 2MASS K$_s$-band magnitudes using the scaling relation from \cite{Wen2013}. In Figure \ref{figure_gas_fraction_neighbors}, although more precise BASS measurements are used for our target host galaxies, the comparison remains valid because the subsamples are divided based on the stellar mass of the neighboring galaxies. For \HI\ masses, we utilize primary beam corrected moment~0 maps to integrate the flux density and convert to \HI\ masses using equation \ref{eq_HI_mass_Horellou01}. For three neighboring galaxies that are caught on the edge of the primary beam (Z~504-097, AGC~123012, and AGC~123016; all in BAT~129 group), we adopted reliable \HI\ flux from HYPERLEDA\footnote{\url{http://atlas.obs-hp.fr/hyperleda/}}. For all neighbors in the NGC~3079 system (BAT~ID~484), we adopted the \HI\ masses from Table 2 of \cite{n3079_WSRT_2015MNRAS}.

In Figure \ref{figure_gas_fraction_neighbors}, our targets are generally found at the high end of \HI\ mass for a given stellar mass, naturally due to our selection criteria. This is also consistent with what \cite{Starikova2011} have shown, i.e., that Chandra/Boötes X-ray AGNs are predominantly located at the centers of dark matter halos, and do not reside as satellites. 
Interestingly, 13 cases are in groups that include at least one neighbor that is gas-poor or severely gas-deficient for a given stellar mass, even after the primary beam correction (see Panels (a) and (b) in Figure \ref{figure_gas_fraction_neighbors}).

However, in 5 cases, all neighboring galaxies are found to be gas-rich as their hosts (see Panel (c) of Figure \ref{figure_gas_fraction_neighbors}). 
Among these 5 systems, three (BAT IDs 64, 83, and 310) have relatively undisturbed \HI\ morphologies, with projected distances to their neighbors of approximately 100~kpc or more. Notably, none of the neighboring galaxies show significant disturbances in their \HI\ distributions. While we cannot completely rule out the possibility that these systems have exchanged gas through their outermost gas disks, they are likely in an early stage of interaction, having experienced no more than a single encounter. In contrast, BAT~1198 and 1202 show more complex cases. Both targets have one \HI-rich neighbor located $\sim$150~kpc away, and are sharing \HI\ envelope with a small nearby galaxy.
These two cases also clearly share gas in velocity space (see Figs. \ref{fig_mom_C_B} and \ref{fig_PVD_appendix}), rather than representing mere superpositions or projection effects.
In particular, in the case of BAT~1202, the kinematically distinct gas clump to the west appears to have an optically unidentified stellar counterpart and is inferred to be supplying gas as it merges with BAT~1202. 

Finally, Panel (d) of Figure \ref{figure_gas_fraction_neighbors} shows the samples with no \HI-detected neighbors. This does not necessarily imply that these examples are located in completely isolated environments or that they have had no interactions with other galaxies. This subsample shows between a few and up to nine nearby objects that may be neighbors, within a 1~Mpc radius and a velocity range of $\pm$500~km~s$^{-1}$.

Even if our AGN hosts have recently gained cool gas from their surroundings, whether the accreted gas will indeed help to activate the central AGN is still questionable. With the scale we are probing in this study, it is hard to make any direct link between the externally driven gas and the AGN activity. Intriguingly, however, among our 22 galaxies, we find that the more luminous samples show more extended \HI\ gas morphologies. This suggests that the gas budget from outside may indeed not be irrelevant to the AGN activity, which will be discussed in the following section. 

\subsection{H{\scriptsize I} and black hole properties}

\begin{figure}[!t]
    \centering
    \includegraphics[width=9.2cm]{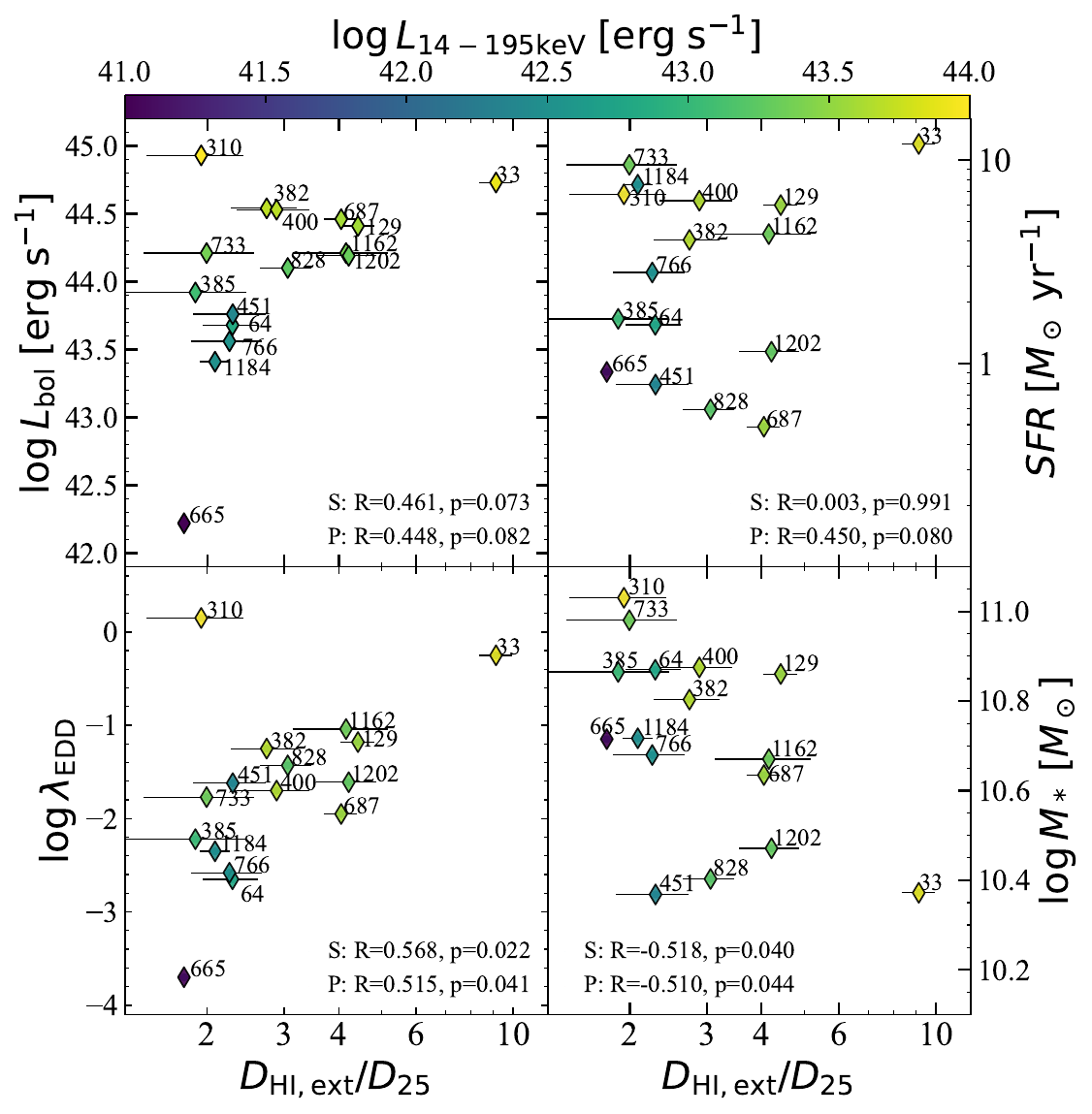}
    \caption{Bolometric luminosity (${L}_{\rm bol}$), Eddington ratio ($\lambda_{\rm Edd}$), star formation rate (SFR), and stellar mass (${M}_*$) as a function of the relative extent of \HI\ gas including low-density gas (${D}_{\rm HI, ext}$) to stellar disk (${D}_{25}$). 
    ${L}_{\rm bol}$ and $\lambda_{\rm Edd}$ are adopted from BASS DR2 \citep{koss2022cat}.
    Each galaxy is colored by the hard X-ray luminosity adopted from \cite{Ricci2017}.
    On each panel, the Spearman rank correlation (S) and the Pearson correlation (P) coefficient, and a p-value between each axis are shown at the bottom.
    There is a marginal, linear correlation between ${L}_{\rm bol}$ and the \HI\ extent in the sense that galaxies with a relatively larger \HI\ disk compared to the stellar disk are more luminous. This also applies to the Eddington ratio, whereas the relative size of \HI\ does not seem to be related to SFR. 
    ${M}_*$ appears to show a marginal negative correlation with ${D}_{\rm HI, ext}/{D}_{25}$, which is naturally expected, as ${D}_{\rm HI, ext}$ is normalized by ${D}_{25}$.
    }
    \label{fig_Lbol_HIextent}
    \centering
\end{figure}

Figure \ref{fig_Lbol_HIextent} shows two AGN-related parameters (${L}_{\rm bol}$, $\lambda_{\rm Edd}$), and two host galaxy-related parameters (SFR and ${M}_*$), as a function of diffuse \HI\ gas extent. The galaxies are colored by their hard X-ray luminosity (${L}_{\rm 14-195keV}$).
As described in Sec \ref{sec: measurements}, here we simply measure the largest, projected extent in the sky at the \HI\ column density that is low enough to present the true gas extent but sufficiently high to be reliable (${N}_{\rm HI} = 5 \times {10}^{19}\text{cm}^{-2}$). 
To robustly measure the extent of diffuse gas around individual galaxies, we used only the data combined with the D-array and with a signal-to-noise ratio greater than 3 at the ${N}_{\rm HI} = 5 \times {10}^{19}\text{cm}^{-2}$ level. As a result, BAT~484 is excluded from the analysis.
We omitted BAT~96 and BAT~1198, which share a large-scale gas envelope with their neighbors at the ${N}_{\rm HI} = 5 \times {10}^{19}\text{cm}^{-2}$ level, and hence hard to measure the extent of individual galaxies. 

Intriguingly, more luminous galaxies, both bolometrically and in hard X-rays, tend to show large \HI\ extents. These luminous systems also have relatively high Eddington ratios, which indicates an active accretion phase. The Spearman rank correlation coefficient for ${L}_{\rm bol}$ and ${L}_{\rm 14-195keV}$ on ${D}_{\rm HI, ext}/{D}_{25}$ is 0.461, with a p-value of 0.073. $\lambda_{\rm Edd}$ has a Spearman coefficient of 0.568 with a p-value of 0.022. 
${M}_*$ shows a Spearman coefficient of -0.518 with a p-value of 0.040. 
None of the other relations (SFR, and ${N}_{\rm HI}$) has a coefficient higher than 0.5. The correlation coefficients are the same for ${L}_{\rm bol}$ and ${L}_{\rm 14-195keV}$ since the Spearman coefficients consider only the ranking of the given data.

Interestingly, when we exclude cases where only \HI-rich neighboring galaxies are present, the correlation becomes more robust (see Figure \ref{fig_Lbol_HIextent_appendix}). The Spearman rank correlation coefficient between ${L}_{\rm bol}$ and ${D}_{\rm HI, ext}/{D}_{25}$ increases to 0.710, with a p-value of 0.007. The strongest correlation is found for $\lambda_{\rm Edd}$, with a Spearman coefficient of 0.824 and a p-value of 0.001, indicating a statistically significant relationship between $\lambda_{\rm Edd}$ and ${D}_{\rm HI, ext}/{D}_{25}$. The correlation with ${M}_*$ remains negative, with a Spearman coefficient of –0.407, indicating an inverse relationship with a p-value of 0.168.

It is remarkable that processes occurring on galactic outskirts or intergalactic scales of several kpc appear to be connected to activity in the central regions on sub-pc scales. This link therefore warrants further detailed investigation and should indeed be tested using the full sample. Nevertheless, when considering the overall properties of our subsample, our best plausible interpretation is as follows.

The observed correlation between the extent of diffuse \HI\ and $\lambda_{\rm Edd}$ could arise if gas accretion - particularly driven by tidal interactions with nearby galaxies for these targets - has been transported to the central regions, thereby fueling the AGN. The fact that this correlation becomes tighter once group (c) systems, which are likely in an early stage of group interaction, are excluded (left panel of Fig. \ref{fig_Lbol_HIextent_appendix}) supports the plausibility of this interpretation.
However, given the large difference in physical scales involved for gas components spanning tens to hundreds of kpc to flow inward to sub-pc scales, the dissipation of angular momentum is indispensable.

While a more detailed investigation of the funneling process is planned as future work, we find that galaxy-scale structures known to play an important role in channeling gas toward the center are quite common in our sample. In the remainder of this section, we therefore review the structures thought to facilitate inward gas transport and examine illustrative examples from our sample.
The first proposed mechanism is galactic-scale structures such as bars or spiral arms \citep[e.g.,][]{Maiolino2000ApJ, OhSree2012ApJS, Chown2019}. Such galactic structures in the disk can facilitate the transfer of angular momentum, aiding the inflow of gas into the central region \citep[e.g.,][]{Audibert2019A&A, Yu2022A&A}. Intriguing galactic structures are indeed observed in some of our samples; seven galaxies (BAT~IDs 310, 382, 400, 484, 733, 1184, and 1198) exhibit stellar bars, while four others (BAT~IDs 33, 63, 129, and 654) show strong optical spiral structures. Among the edge-on galaxies, some targets (e.g., BAT~IDs 129 and 484) display non-regularly rotating \HI\ components, possibly due to streaming motions driven by large-scale gaseous bars \citep[e.g.,][]{Combes2014}.

Alternatively, externally driven forces may help to funnel the gas to the central region. Tidal torque is generally known to increase the angular momentum of galaxies \citep[e.g.,][]{Peebles1969, Ryden1988ApJ}. However, tidal instability also may lead to the formation of a galactic bar \citep[e.g.,][]{Lokas2016ApJ_bar}, which then transports gas inwards as demonstrated by simulations \citep[e.g.,][]{Blumenthal2018MNRAS}.

Major merging can also disturb the gas kinematics, leading to the loss of angular momentum and driving the medium to fall into the SMBH \citep[e.g.,][]{Springel2005, Hopkins2008_coevolution_1, Koss2010, Gaspari2013, Ellison2019, Secrest2020}. Based on their relatively unperturbed stellar disks, however, none of our targets seems to be undergoing major merging, even though the majority of our sample is found with neighboring galaxies. Furthermore, our targets are not particularly the hosts of severely obscured AGNs based on $N_{\rm H}$ from X-ray spectra, indicating that they are indeed not in the middle of a merging event \citep{Ricci2021_GOALS}. Our sample is also moderately luminous in X-ray (${L}_{\rm X}\sim 10^{42-44}\text{erg}\;\text{s}^{-1}$), unlike merging systems which can often be more luminous in X-ray \citep[e.g.,][]{Treister2012, Glikman2015, Hong2015, KimMJ2021ApJS}.

As an alternative externally driven impact, \cite{Poggianti2017Natur} suggested that ram pressure stripping by intracluster medium (ICM), with thermal instabilities and turbulent motions, is capable of funneling the gas towards the galaxy center. They argued that this process can help gas accretion onto the central BH and trigger the activity. But none of our sample resides in dense environments like galaxy clusters, implying that the ICM is unlikely to play any role in our targets.

Based on all these, our galaxies are likely the cases where cool gas might have flowed in from the surroundings, particularly through tidal interactions with neighbors, as argued in \ref{discuss:local_env}. Then it could have been funneled into the central region by galactic bars and/or strong spiral arms, reaching the central black hole.
\citet{Koss2011} also support this, finding higher fractions of massive spirals with well-developed arms and mergers (see also \citeauthor{Koss2018} \citeyear{Koss2018}, who identify an excess of kpc scale mergers in many obscured AGNs) among BAT AGN hosts compared to the comparison sample of inactive galaxies. Likewise, a recent analysis of $>$1000 BAT AGN hosts \citep{BASS_Miguel_Parra_morph_2025} finds an elevated incidence of bars and disturbed optical morphologies compared to control galaxies, which also supports how galactic structures could aid gas funneled into the central AGN.

It is also important to understand how accreted gas entering the center of the host galaxy transforms into a denser form of gas.
Among our sample of 22, however, single-dish CO data are available for only nine targets from \cite{Koss2021}, and no clear trends in molecular gas are observed among those galaxies. To better understand the link between large-scale gas supply and AGN activity, resolved molecular gas properties will have to be studied for our large \HI\ sample of AGN hosts.

\vspace{5mm}

\section{Conclusions and future work} \label{sec: conclusion}

We have probed the resolved \HI\ gas properties of 22 BASS hard X-ray-selected AGN hosts. The sample is a good representative of \HI-rich Seyfert galaxies (low- to moderate-luminosity X-ray-selected AGNs) in the Local Universe. Apart from one common feature - a large \HI\ gas reservoir - the targets display a range of characteristics, including differences in stellar morphology and star formation activity. 

Our \HI-rich AGN hosts exhibit various properties in atomic gas morphology and kinematics. 
Some galaxies show hints of tidal interactions with their surroundings, such as extended \HI\ tails or common gas envelopes.
This indicates that our \HI-rich AGN hosts are primarily located in an environment where gas exchange with neighbors can occur relatively easily. This is also consistent with the previous finding that X-ray AGNs are preferentially found in group environments. In particular, the majority of our targets are the most massive systems in their respective groups, making our sample likely to gain gas rather than to be gas donors.

We find that the extent of diffuse atomic gas correlates with the bolometric/X-ray luminosities and the Eddington ratio. Given that the extended, diffuse \HI\ gas can be a footprint of the gas accretion from the surroundings, our finding suggests that the externally driven gas may be a key gas supply in \HI-rich, low- to moderate-luminosity X-ray-selected AGN hosts.  
{This relation must be further probed with a larger sample, for which statistically reliable conclusions can be drawn.}

Although we find that the surrounding environment could be a source of gas supply for these particular \HI-rich AGN hosts, some of the galaxies are located in relatively isolated environments. Therefore, it is necessary to examine cases with more diverse \HI\ masses, as not all AGN hosts potentially have gas donors nearby. 
Despite some hints for the link between possibly externally driven cool gas and AGN activity, further details of how the accreted gas is transferred to the center, turning on the central AGN, are still an open question. To tackle this, multi-wavelength studies on various scales will be essential, which is one of the key goals of the BASS project.

Lastly, we are currently analyzing \HI\ gas and radio continuum data of $\sim$100 hard X-ray-selected BASS AGN hosts. Soon, we hope to provide more insights on the role of neutral hydrogen in star formation as well as AGN activity with a large sample spanning a broader range of properties. 

\vspace{5mm}

\begin{acknowledgements}
We are grateful to N.Shafi and T.Oosterloo for kindly providing the WSRT \HI\ cube of NGC~3079 for this work.
The National Radio Astronomy Observatory is a facility of the National Science Foundation operated under cooperative agreement by Associated Universities, Inc.
This research was supported by the Korea Astronomy and Space Science Institute under the R\&D program(Project No. 2025-9-844-00) supervised by the Korea AeroSpace Administration.
A.C. and J.K. acknowledge support by the NRF, grant Nos. RS-2022-NR069020 and RS-2022-NR070872. J.K. also acknowledges support by the BK21 Research and Studies Overseas Educational Program.
M.J.K. acknowledges support from NASA through ADAP award 80NSSC22K1126. 
KO acknowledges support from the Korea Astronomy and Space Science Institute under the R\&D program (Project No. 2025-1-831-01), supervised by the Korea AeroSpace Administration, and the National Research Foundation of Korea (NRF) grant funded by the Korea government (MSIT) (RS-2025-00553982).
F.E.B. acknowledges support from ANID-Chile BASAL CATA FB210003, FONDECYT Regular 1241005,
and Millennium Science Initiative, AIM23-0001.
Y.D. acknowledges the financial support from a Fondecyt postdoctoral fellowship (3230310).
M.K. was supported by the National Research Foundation of Korea (NRF) grant funded by the Korean government (MSIT) (No. RS-2024-00347548).
I.M.C. acknowledges support from ANID programme FONDECYT Postdoctorado 3230653.
C.R. acknowledges support from Fondecyt Regular grant 1230345, ANID BASAL project FB210003, and the China-Chile joint research fund.
M.S. acknowledges financial support from the Italian Ministry for University and Research, through the grant PNRR-M4C2-I1.1-PRIN 2022-PE9-SEAWIND: Super-Eddington Accretion: Wind, INflow and Disk-F53D23001250006-NextGenerationEU.
B.T. acknowledges support from the European Research Council (ERC) under the European Union's Horizon 2020 research and innovation program (grant agreement number 950533) and from the Israel Science Foundation (grant number 1849/19).
We acknowledge the usage of the HyperLeda database (\url{http://atlas.obs-hp.fr/hyperleda/}).
This research has made use of the Astrophysics Data System, funded by NASA under Cooperative Agreement 80NSSC21M00561.
\end{acknowledgements}

\bibliographystyle{aa}
\bibliography{new25}

\onecolumn
\begin{appendix}
\section{\HI\ moments maps for the rest of the samples}\label{appendix:continued_mom_maps}
\FloatBarrier

\begin{figure*}[htb]
    \centering
    \includegraphics[width=12.4cm]{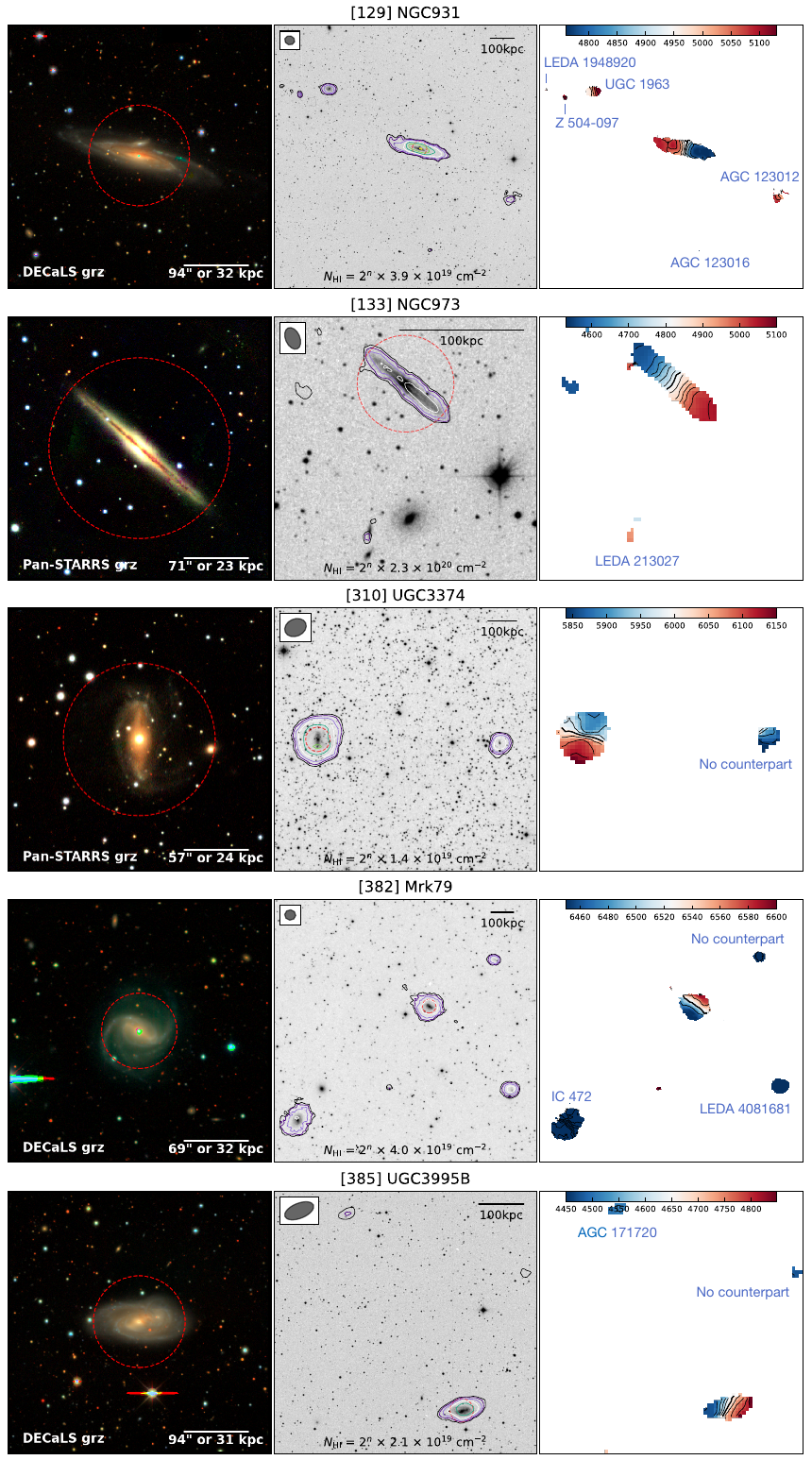}
    \caption{Figure \ref{fig_mom_C_A} continued.}  
    \label{fig_mom_C_B}
\end{figure*}

\begin{figure*}[pth]
    \centering
    \ContinuedFloat
    \subfloat[]{\includegraphics[width=12.4cm]{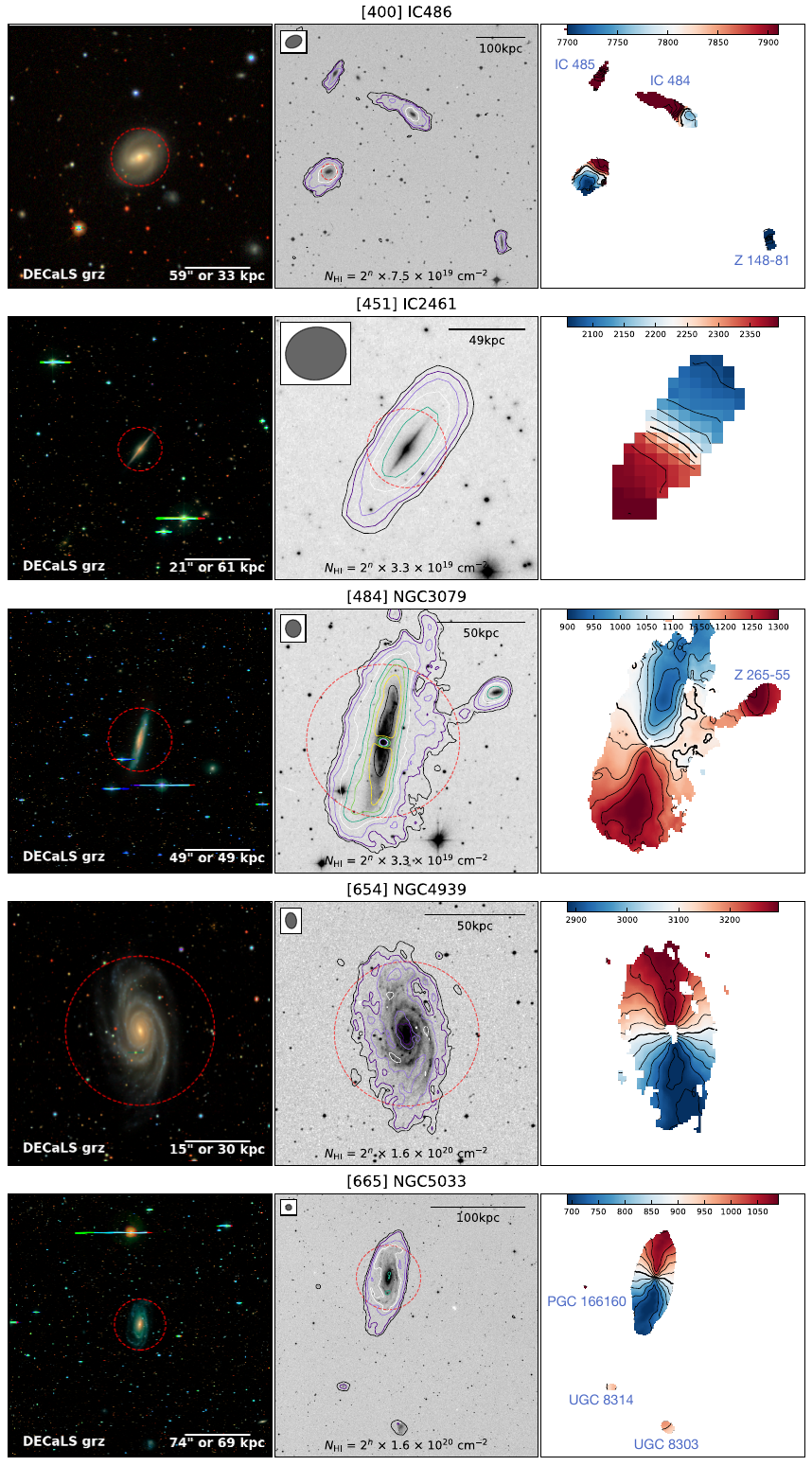}}
    \caption{continued.}
\end{figure*}

\begin{figure*}[pth]
    \centering
    \ContinuedFloat
    \subfloat[]{\includegraphics[width=12.4cm]{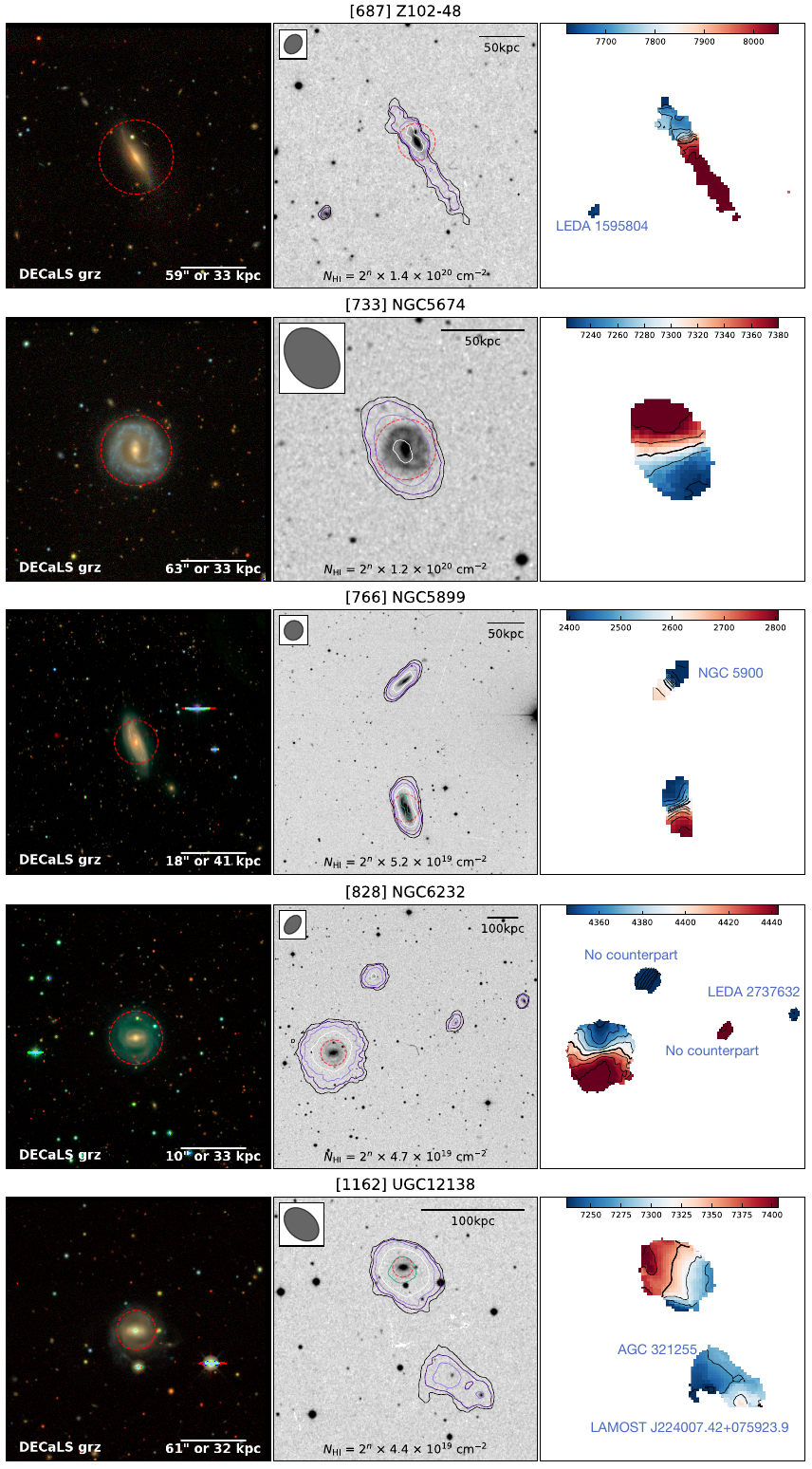}}
    \caption{continued.}
\end{figure*}

\FloatBarrier
\begin{figure*}[pth]
\centering
\ContinuedFloat
\includegraphics[width=11.9cm]{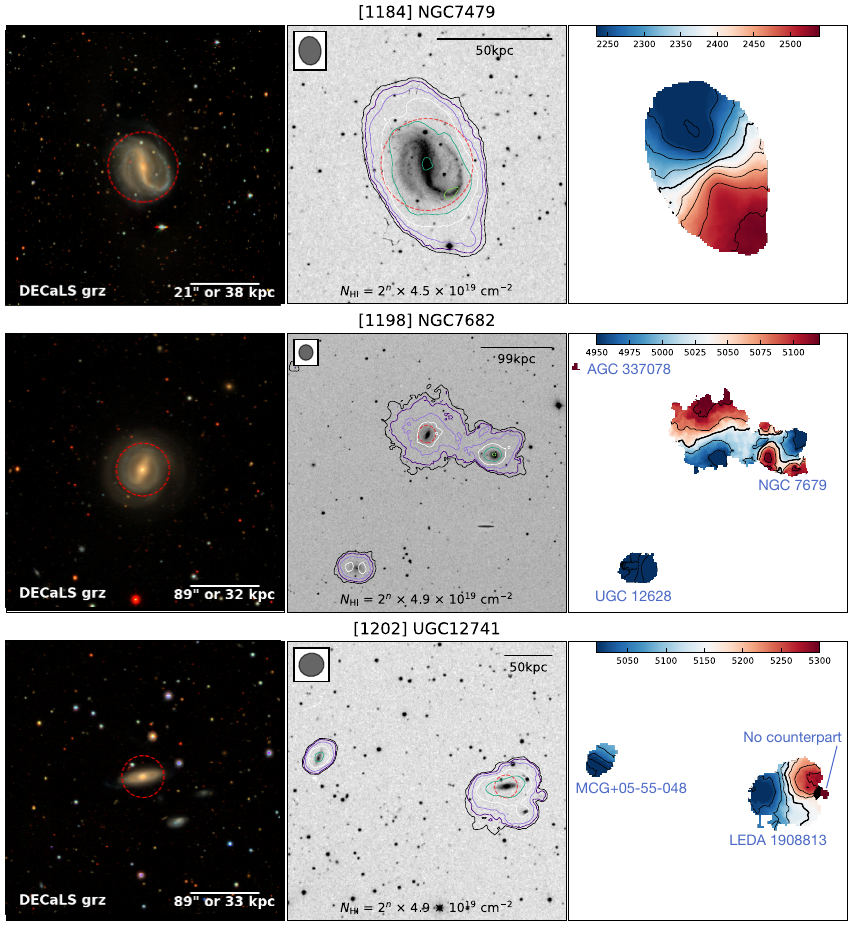}
\caption{\footnotesize continued.}
\end{figure*}

\FloatBarrier
\vspace{-5mm}
\section{Nearby companions sharing \HI\ gas envelopes}
\noindent

\vspace{-5mm}

\begin{figure}[hb]
    \centering
    \includegraphics[width=16.5cm]{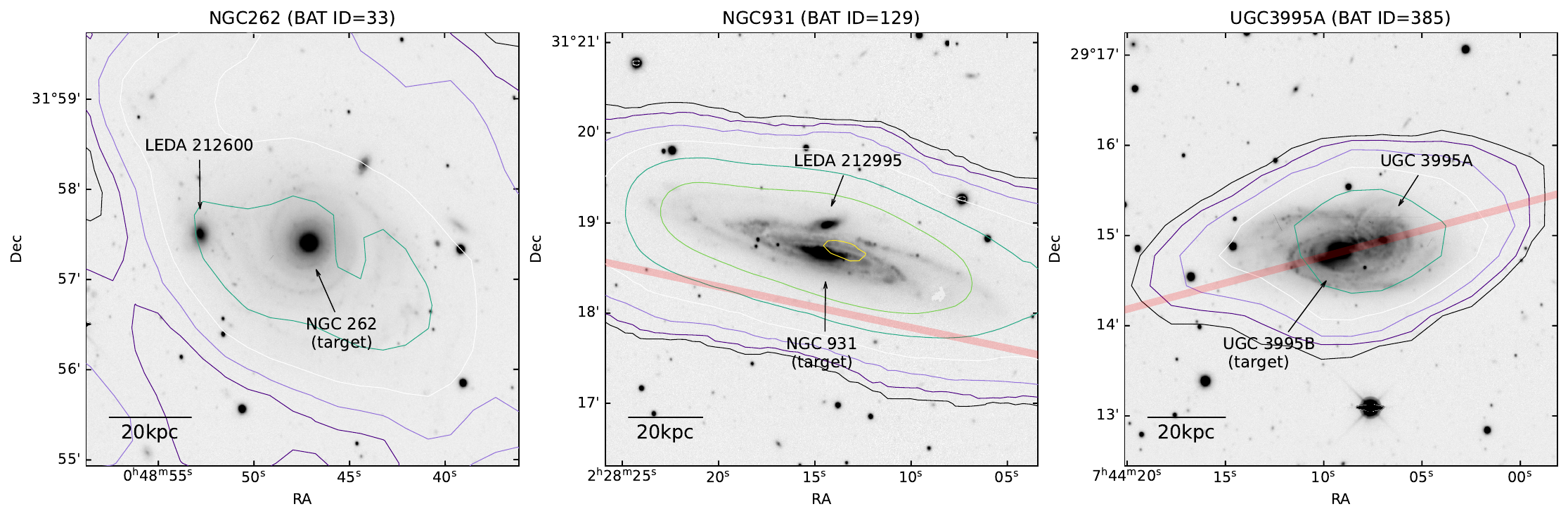}
    \caption{
    DECaLS r-band images of redshift-confirmed close galaxy companions sharing common \HI\ gas envelopes with our targets. The \HI\ contours are identical to those presented in Figures \ref{fig_mom_C_A} and \ref{fig_mom_C_B}. The red lines mark the locations of the position–velocity diagram (PVD) cuts displayed in Figure \ref{fig_PVD_appendix}. NGC~262 and its companion LEDA~212600 are considered to form an interacting but largely undisturbed system \citep[e.g.,][]{Simkin1987, Anton2002}, with a projected separation of 22~kpc \citep[][]{Koss2010}.
    NGC~931 and LEDA~212995 constitute a pair with a projected separation of $\sim6$~kpc \citep[][]{Koss2012dualAGN}. Although the spatial resolution of the \HI\ beam is insufficient to fully resolve the system, the PVD reveals a gas component distinct from the main \HI\ disk (see Figure \ref{fig_PVD_appendix}). 
    UGC~3995A and UGC~3995B are a close interacting galaxy pair separated by $\sim50$km~s$^{-1}$ in radial velocity \citep[][]{Holwerda2013}. However, the absence of strong tidal distortions suggests that the system is likely in an early stage of interaction \citep[e.g.,][]{Marziani1999}.
    }
\label{figure_companion}
\end{figure}

\FloatBarrier
\section{\HI\ Position-Velocity Diagrams}

\begin{figure*}[hbt]
    \centering
    \includegraphics[width=18cm]{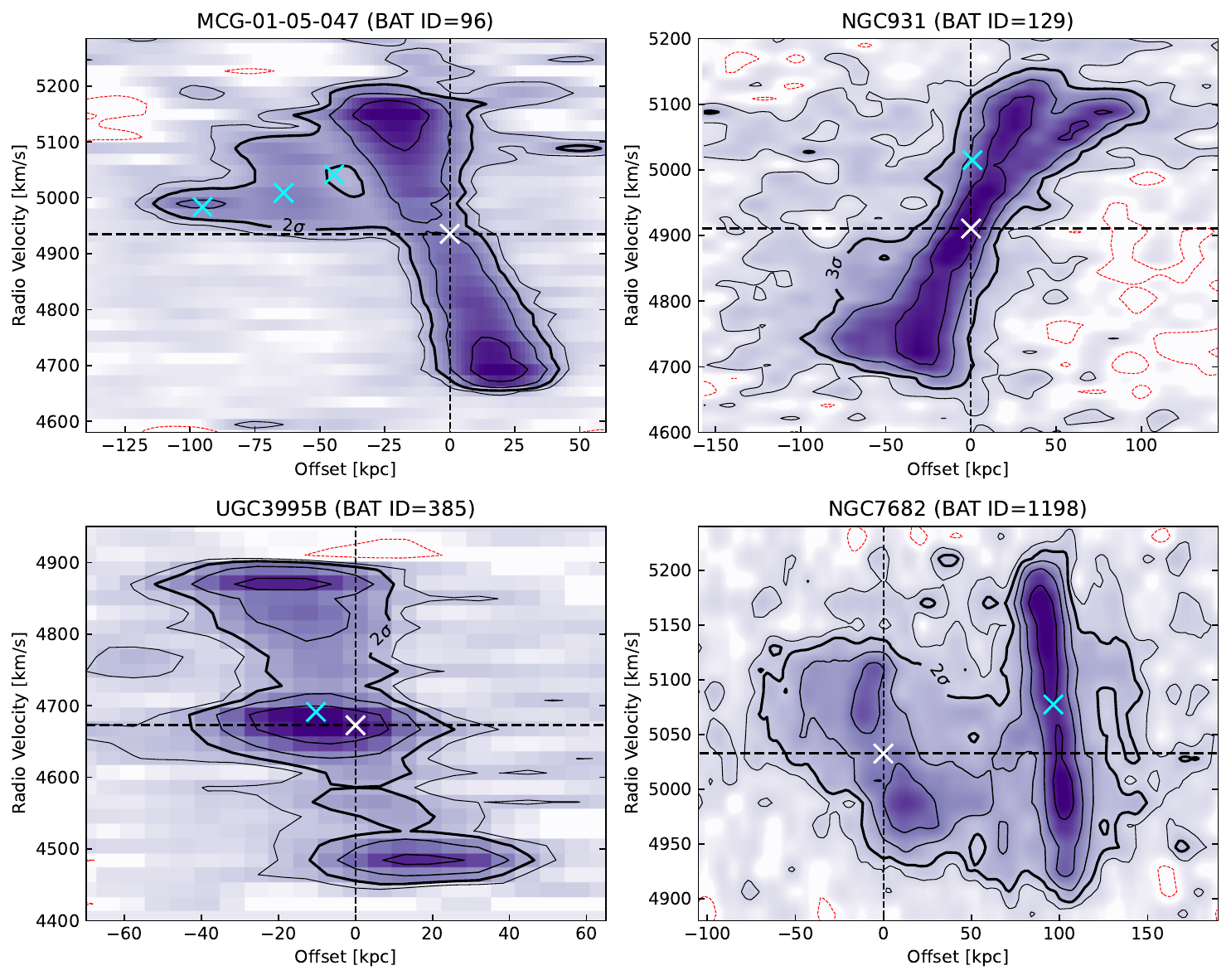}
    \caption{\HI\ PVDs for BAT~IDs 96, 129, 385, and 1198. The outer contour levels are [-1$\sigma$, 1$\sigma$], where negative lines are shown in dotted red lines.
    The inner contour levels are $2^n\times \sigma$, where n=2,3,4, and so on, with the 2$\sigma$ or 3$\sigma$ level indicated by a thick contour in each plot.
    {The velocities and projected offsets of neighboring galaxies, derived as described in the text, are overlaid as colored crosses in each panel.}
    In MCG-01-05-047, a diffuse \HI\ envelope appears to connect the galaxy with its two neighbors, both spatially and in the receding velocity field. For NGC~931, a gas component deviating from the main disk kinematics is present in the approaching-velocity channels. Although its precise location cannot be resolved with the current \HI\ beam, it is likely a signature of external gas inflow \citep[see also Figure 19 of][]{Kuo2008}. In the case of UGC~3995, the \HI\ PVD shows three peaks. The components are not spatially resolved because the two galaxies form a tight pair (see Figure \ref{figure_companion}). However, as shown in Figure 34 of \citeauthor{Kuo2008} \citeyear{Kuo2008}, one gas disk appears in the approaching channels while the other becomes visible in the receding channels. Finally, NGC~7682 and NGC~7679 clearly share a common \HI\ envelope that spans a velocity range of $\sim$150~km~s$^{-1}$ \citep[see Figure 4 of][for a channel map]{Kuo2008}.
    }
\label{fig_PVD_appendix}
\end{figure*}

Figure \ref{fig_PVD_appendix} presents \HI\ position–velocity diagrams (PVDs) for four galaxies in order to more clearly illustrate possible gas connections with neighboring systems.
The PVD cut positions are displayed in Figure \ref{figure_companion} and were selected to best reveal kinematic \HI\ connections with neighboring galaxies. 
For MCG-01-05-047, the PVD slice was aligned along the major axis of the target galaxy.
For NGC~7682, the PVD slice was chosen to pass through the centers of both NGC~7682 and its companion NGC~7679.
For each target, the optical center of the host galaxy is adopted as the center of the PVD slice. The systemic velocity ($v_{\rm sys}$) in Table \ref{tab:measured_properties} and the reference position is marked as a white cross in Figure \ref{fig_PVD_appendix}.

Neighboring galaxy positions are projected onto the PVD slice and measured as offsets from the slice center. 
In the case of MCG-01-05-047 system, the velocities of the two known neighboring galaxies, LEDA~1071790 and LEDA~1071479, are taken from 
\cite{Spekkens2005AJ....129.2119S} and \cite{Jones2009MNRAS.399..683J_6dF}, respectively. In one case where no optical counterpart is known, the velocity of the \HI\ gas component measured at the optical center is adopted instead.
For the remaining systems, the velocities of the neighboring galaxies LEDA~212995, UGC~3995A, and NGC~7679 are adopted from \cite{n931_and_companion_1992A&A_Amram}, \cite{Marziani1999}, and \cite{Buson2006A&A...447..441B}, respectively. Velocities reported in the optical definition are converted to the radio velocity. The resulting positions are marked as cyan crosses on the PVDs in Figure \ref{fig_PVD_appendix}.

\FloatBarrier
\section{\HI\ extent correlation in subsamples}
\noindent

\begin{figure}[hbt]
    \centering
    \includegraphics[width=\textwidth]{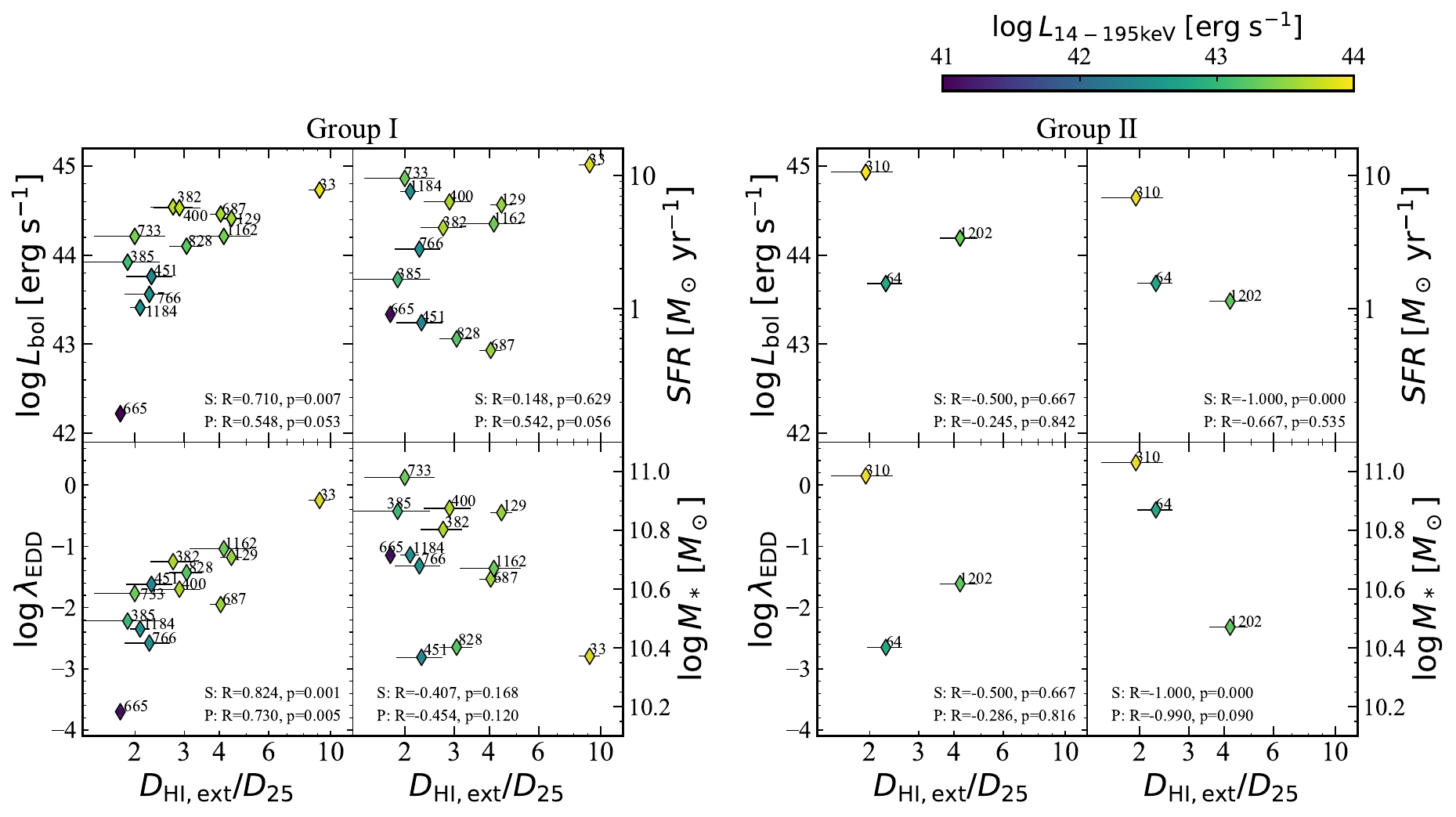}
    \caption{Same as Figure \ref{fig_Lbol_HIextent} divided into subsamples.
    (left) \textbf{Group I}: Subgroup with at least one gas-poor neighbors, or neighbor(s) with no \HI. (Panel (a), (b), and (d) in Figure \ref{figure_gas_fraction_neighbors}). 
    (right) \textbf{Group II}: Subsample with only gas-rich neighbors (Panel (c) in Figure \ref{figure_gas_fraction_neighbors}).
    Group I systems are showing tighter correlation of Bolometric luminosity (${L}_{\rm bol}$), Eddington ratio ($\lambda_{\rm Edd}$) as a function of the relative extent of \HI\ gas including low-density gas (${D}_{\rm HI, ext}$) to stellar disk (${D}_{25}$). 
    Hence, Group I systems are more likely to have exchanged gas with their surroundings, compared to the Group II shown in the right panel, where these systems are likely in an earlier phase of group interaction.}
\label{fig_Lbol_HIextent_appendix}
\end{figure}

\twocolumn
\end{appendix}

\end{document}